\documentclass[preprint,nofootinbib,showpacs,preprintnumbers,amsmath,amssymb]{revtex4}

\usepackage{graphicx}
\usepackage{dcolumn}
\usepackage{bm}

\begin{document}


\title{The Matsubara-Fradkin Thermodynamical Quantization of Podolsky Electrodynamics}

\author{C. A. Bonin}
 \email{bonin@ift.unesp.br}
 \author{B. M. Pimentel}%
 \email{pimentel@ift.unesp.br}
 \affiliation{%
Instituto de F\'{i}sica Te\'{o}rica UNESP - S\~{a}o Paulo State University.\\
Caixa Postal 70532-2, 01156-970 S\~{a}o Paulo, SP, Brazil.
}%

\date{\today}

\begin{abstract}
In this work we apply the Matsubara-Fradkin formalism and the Nakanishi's auxiliary field method to the quantization of the Podolsky electrodynamics in thermodynamic equilibrium. This approach allows us to write consistently the path integral representation for the partition function of gauge theories in a simple manner. Furthermore, we find the Dyson-Schwinger-Fradkin equations and the Ward-Fradkin-Takahashi identities for the Podolsky theory. We also write the most general form for the polarization tensor in thermodynamic equilibrium.
\end{abstract}

\maketitle

\section{Introduction}

There are several ways to take finite-temperature effects into account in a quantum theory. For instance, we can start with the path integral representation for the partition function of the system \cite{Kapusta}. This representation is based on an analogy between the partition function and the vacuum-to-vacuum transition amplitude and it is called imaginary-time formalism. Once the partition function is evaluated, all thermodynamical quantities can be computed from it. The downside of this technique appears most significantly when we consider constrained systems. In those cases we usually tread a long path sorting out the true physical degrees of freedom of the theory. For path integrals, this program was developed by Senjanovic \cite{Senjanovic}. Since all gauge theories are constrained, this step cannot be avoided in this approach. Instead of this method, we may consider the real-time formalism, in which the time parameter of the field operators becomes a complex number \cite{Das, Le Bellac}. This method is very suitable to study small deviations from equilibrium. Also, in this formalism a natural separation between quantities that depend on the temperature and quantities that do not occurs. However, for coherence of the method, the Green functions of the system become matrices, with each entry associated with a segment of a path in the complex-time plane. The approach known as \textit{Thermo Field Dynamics} conceives time-evolution of thermal operators as well \cite{Umezawa, Umezawa advanced}. In this formulation, thermal averages of operators are written as expectation values of these operators in the so-called \textit{thermal vacuum}. However, due to an operation called \textit{tilde operation}, the number of the fields must be doubled. As an alternative to these methods, we can begin by defining our ensemble in thermodynamic equilibrium through its density matrix. This last formulation was the one designed by Matsubara for non-relativistic systems \cite{Matsubara}. In its original form, Matsubara's approach only works for systems described by the canonical ensemble. Soon after, Fradkin extended Matsubara formalism to include also the grand-canonical ensemble and, therefore, he developed a suitable technique to deal with (relativistic) quantum field theories, where the number of particles is not fixed \cite{Fradkin1,Fradkin2}. Fradkin also used Schwinger's source fields in order to make the computation of ensemble averages easier. Among the interesting features of the ``Matsubara-Fradkin" formalism, we mention the following. All positivist quantum theories \textit{assume} that the state of the system is described by a vector in the Hilbert space or, more generally, by a \textit{density matrix}. In other words, it is a postulate of positivist quantum theories that the density matrix of the system contains \textit{all the information about the system}. So, from the physical point of view, it is more natural and fundamental to start from the density matrix. Furthermore, in this method we can find a path integral representation for the partition function that is based on no analogy. Also,  it is possible to extend the formalism to include time-dependent Green functions, describing, in this way, small deviations from equilibrium in much the same way the real-time formalism does \cite{Fradkin2}. As a final remark, we point out that the spacetime's Euclidean structure for systems in equilibrium emerges naturally from the formalism. As a consequence, there is a mathematical correspondence between expressions for zero-temperature theories and for ``hot fields". So, the Matsubara-Fradkin approach to thermodynamical quantization seems a reasonable candidate when compared with the other, far more common, techniques. However, the Matsubara-Fradkin formalism does not deal, by itself, with the question of gauge theories: the quantization of a gauge theory is always a challenge. Even if we consider the simplest of all gauge theories, \textit{viz.} the ordinary electrodynamics, we still face some problems in its quantization process. We can take two major alternative routes: we can follow Dirac's footsteps or Nakanishi's. In the former case, all the constraints are dealt with in the classical level, and we have simple rules relating classical Dirac brackets with commutators and anti-commutators for the quantum theory \cite{Dirac}. The main drawback of this method is that the Lorentz covariance of the theory is explicitly broken. On the other hand, if we follow Nakanishi's quantization procedure, we start with a quantum Lagrangian density which incorporates a particular gauge choice \textit{ab initio} \cite{Nakanishi paper, Nakanishi}. Due to that choice, the theory ceases to be gauge invariant. Despite this explicit breaking of the gauge symmetry, Nakanishi's method has one main advantage: it is Lorentz covariant. When we consider a gauge theory in thermodynamic equilibrium, there is an advantage in a quantization procedure that maintains Lorentz covariance in all steps. If we perform the Dirac's constraints analysis, in some stage we usually pass from a non-covariant gauge choice to a covariant one. This is accomplished using, for example, the Faddeev-Popov trick in the path integral representation. By comparison, we see that the Nakanishi's quantization method plays a simpler role, being Lorentz covariant from the beginning to the end. If we join together Nakanishi's method with the Matsubara-Fradkin formalism we are able to write down the correct path integral representations for partition functions of gauge theories \textit{without} the need of performing Senjanovic's procedure for constrained theories. In the present paper we deal with the question of quantizing a generalization of the Maxwell electrodynamics in thermodynamic equilibrium. Such a generalization is known as \textit{Podolsky electrodynamics}.

In 1942, Podolsky proposed a theory for the electromagnetic interaction as an alternative to Maxwell's \cite{Podolsky_1}. Podolsky theory treats the action for the electromagnetic interaction as a functional depending not only on the electromagnetic field and its first derivatives, but also on its second-order derivatives. As a consequence, it is possible to split the Podolsky field in a sum of a massless vector field with a massive one. Furthermore, Podolsky theory depends on a free parameter. This parameter can be interpreted as the mass of the massive vector field. Despite the presence of the massive field, Podolsky theory is gauge, as well as Lorentz, invariant. As a matter of fact, Cuzinatto, de Melo, and Pompeia proved in 2007 that Podolsky theory is the only possible extension (regarding theories with second-order derivatives) for Maxwell electromagnetism that maintains these two symmetries in the Abelian case \cite{Cuzinatto_1}. Besides, Podolsky electrodynamics can be experimentally tested. In \cite{Bonin} we, along with co-workers, studied the free Podolsky field in thermodynamic equilibrium using the imaginary-time formalism. In that work we showed that a correction to the Stefan-Boltzmann law is expected if the Podolsky field is the underlying field in the electromagnetism. Besides, we have used data from cosmic microwave background radiation to set a thermodynamical limit to the free parameter of the theory.  In \cite{Cuzzinatto_2}, Cuzinatto \textit{et al.} proposed several other experiments for measuring the Podolsky parameter. Moreover, in \cite{BPZ}, one of us with collaborators has shown that the electron self-energy and the vertex function of Podolsky theory at zero temperature are both free from ultraviolet divergences in the lowest-order of perturbation theory. This does not happen in Maxwellian QED.  Our goal in this paper is to study Podolsky quantum electrodynamics, also called Generalized QED$_4$ (GQED$_4$), in thermodynamic equilibrium. We start with a quick review about the Matsubara-Fradkin formalism.  Using widely known results from the ordinary QED$_4$, we write down the field equations for the fermionic sector of the GQED$_4$. For the gauge sector, instead, we use the more convenient Nakanishi's procedure. Once we have all  field equations, we find the set of equations satisfied by the thermodynamical generating functional of Podoslky theory. From the generating functional we obtain the Dyson-Schwinger-Fradkin equations and the Ward-Fradkin-Takahashi identities.

This paper is organized as follows. In section \ref{MF formalism} we review the Matsubara-Fradkin formalism for a general quantum field theory. In section \ref{GQED} we study the GQED$_4$ in thermodynamic equilibrium using the Matsubara-Fradkin formalism. In section \ref{Green} we study ensemble averages of some special operators. In section \ref{sec partition function} we find path integral representations for the thermodynamical generating functional and for the partition function, where we show that the Podolsky free partition function coincides with the one we have found in our previous work \textit{via} imaginary-time technique. The Dyson-Schwinger-Fradkin equations for the GQED$_4$ in thermal equilibrium are found in section \ref{DSF}.  In section \ref{WFTsection} we seek the Ward-Fradkin-Takahashi identities for the theory in thermodynamic equilibrium. We reserve section \ref{final remarks} for our final remarks. As long as Minkowski spacetime is concern, we use the metric signature $\left(+,-,-, ..., -\right)$.

\section{The Matsubara-Fradkin formalism}\label{MF formalism}

In this section we review the Matsubara-Fradkin formalism. As far as we known, although Fradkin extensively used this formalism at finite temperature, he only carefully demonstrated its zero-temperature version \cite{Fradkin zero temperature}. For this reason, we present his formalism in thermal equilibrium from the beginning.

The density matrix contains all the information about the physical ensemble of the quantum system.
For a system in thermodynamic equilibrium the grand-canonical ensemble density matrix reads\footnote{The density matrix also depends explicitly on all the chemical potentials and implicitly on the physical (hyper)volume $V$. We denote it by $\widehat{\rho }\left( \beta \right)$ just for the sake of notation. This comment also holds for the partition function (\ref{partition}).}

\begin{equation}
\widehat{\rho }\left( \beta \right) =\exp \left[ -\beta \left( \widehat{H}%
-\mu _{j}\widehat{N}_{j}\right) \right] ,  \label{rho}
\end{equation}
where $\widehat{H}$ is the  Hamiltonian and $\left\{ \widehat{N}_{j}\right\} $ is the set of all Noether charges associated with continuous internal symmetries of the theory. Each of these charge operators is conserved. We assume implicit sum in the index $j$ in this expression over all Noether charges. $\beta $ is the inverse of the thermal energy and $\left\{ \mu _{j}\right\} $ is the set of chemical potentials related with the Noether charges.  We assume $\widehat{H}$ depends on local fields which we denote by $ \widehat{\phi }_{j}\left( \mathbf{x}\right)$. Since stationarity is a requirement for thermodynamic equilibrium, these fields are time-independent. In this section we also assume that our spacetime is $\left(D+1\right)$ - dimensional. It is important to stress that the Hamiltonian $\widehat{H}$  takes into account also \textit{all the interactions} among the fields.

The grand-canonical density matrix satisfies the Bloch equation,
\begin{equation}
\frac{\partial \widehat{\rho }\left( \beta \right) }{\partial \beta } =-\left( \widehat{H}-\mu _{j}\widehat{N}%
_{j}\right) \widehat{\rho }\left( \beta \right),\label{bloch equation}
\end{equation}
and its trace is the partition function:

\begin{align}
Z\left(\beta\right)\equiv\mbox{Tr}\left[\widehat{\rho }\left( \beta \right)\right].\label{partition}
\end{align}

We also define the thermal average of an arbitrary operator $\widehat{F}$ as

\begin{equation}
\left\langle\widehat{F}\right\rangle\equiv\frac{\mbox{Tr}\left[\widehat{\rho }\left(\beta\right)\widehat{F}\right]}{Z\left(\beta\right)}.\label{average}
\end{equation}

In order to make it easier for us to compute thermal averages we consider the presence of classical external sources $ s_{j}$
associated with the several fields $ \widehat{\phi }_{j}$. The system is, therefore, described by the \textit{total Hamiltonian} $\widehat{H}_{T}$:

\begin{equation}
\widehat{H}_{T}=\widehat{H}+\widehat{H}_{s},
\end{equation}%
where $\widehat{H}_{s}$ is the \textit{source Hamiltonian}. We write it as
$\widehat{H}_{s}=\int \widehat{\mathcal{H}}_{s}\left( \mathbf{%
x}\right) d^{D}x$, with the source Hamiltonian density $\widehat{\mathcal{H}}_{s}$ given by

\begin{equation}
\widehat{\mathcal{H}}_{s}\left( \mathbf{x}\right) =-\frac{1}{2}\sum_{j=1}^{\overline{n}}\left[
s_{j}\left( \mathbf{x}\right) \widehat{\phi }_{j}\left( \mathbf{x}\right)
-\left(-1\right)^{P_j}\widehat{\phi }_{j}\left( \mathbf{x}\right)
s_{j}\left( \mathbf{x}\right) \right] .
\end{equation}

Here, $P_j=1$ if the source $s_{j}$ commutes with its associated field $\widehat{\phi }_{j}\left( \mathbf{x}\right)$ and $P_j=2$ if these objects anti-commutes with each other. $\overline{n}$ is the number of fields of the problem. Due the presence of the sources, the density matrix now depends on the total Hamiltonian:
\begin{equation}
\widehat{\rho }_{s}\left( \beta \right) =\exp \left[ -\beta \left( \widehat{H%
}_{T}-\mu _{j}\widehat{N}_{j}\right) \right] .\label{rho s}
\end{equation}

We can check that this density matrix satisfies the Bloch equation (\ref{bloch equation}) with $\widehat{H}$ replaced by $\widehat{H}_T$.
 Since for the great majority of physical problems the fields do not commute with $\widehat{H}-\mu _{j}\widehat{N}_{j}$, we have $\widehat{\rho}_{s}\left(\beta \right)\neq\widehat{\rho}\left(\beta\right)\exp\left(-\beta\widehat{H}_{s}\right) $. In order to write $\widehat{\rho}_{s}\left(\beta \right)$ in a more convenient form, we try the following \textit{Ans\"{a}tz}:

\begin{equation}
\widehat{\rho }_{s}\left( \beta \right) =\widehat{\rho }\left( \beta \right)
\widehat{S}\left( \beta \right).  \label{rho em termos de rho zero}
\end{equation}

Deriving this \textit{Ans\"{a}tz} with respect to $\beta$ and making use of the Bloch equations, we find:

\begin{equation}
\frac{\partial \widehat{S}\left( \beta \right) }{\partial \beta }=-\widehat{H%
}_{s}\left( \beta \right) \widehat{S}\left( \beta \right)  \label{S}
\end{equation}

Here, we have defined for any operator $\widehat{F}$ its dependence with the temperature in the ensemble without external sources:

\begin{equation}
\widehat{F}\left( \tau \right) \equiv \widehat{\rho }^{-1}\left( \tau
\right) \widehat{F}\,\widehat{\rho }\left( \tau \right) .\label{without}
\end{equation}

Integrating (\ref{S}) and using (\ref{rho}), (\ref{rho s}), and (\ref{rho em termos de rho zero}), we get the quantum Volterra equation:

\begin{equation}
\widehat{S}\left( \beta \right) =\widehat{1}-\int_{0}^{\beta }d\tau \widehat{%
H}_{s}\left( \tau \right) \widehat{S}\left( \tau \right) .
\label{Volterravol}
\end{equation}

Using an iteration technique, we can find the solution for this equation:

\begin{equation}
\widehat{S}\left( \beta \right) =T\left\{ \exp \left[ -\int_{0}^{\beta }d\tau \widehat{H}_{s}\left( \tau
\right) \right] \right\} .
\end{equation}

The ordering operator for two fields is $T\left[ \widehat{A}\left( \tau\right) \widehat{B}\left( \tau
\right) \right]=\widehat{A}\left( \tau\right) \widehat{B}\left( \tau
\right)$ and\footnote{The Heaviside step function $\theta(\tau)$ is equal to 1 if $\tau\geq 0$ and equal to zero if $\tau<0$.}

\begin{align}
T\left[ \widehat{A}\left( \tau _{1}\right) \widehat{B}\left( \tau
_{2}\right) \right] \equiv&\,
\theta \left( \tau _{1}-\tau _{2}\right) \widehat{A}\left( \tau _{1}\right)
\widehat{B}\left( \tau _{2}\right) \pm \theta \left( \tau _{2}-\tau
_{1}\right) \widehat{B}\left( \tau _{2}\right) \widehat{A}\left( \tau
_{1}\right) \label{lamel}
\end{align}
if $\tau_1\neq\tau_2$. The minus sign is used when both operators are Grassmannian and the plus sign otherwise. We define a generalization of the operator $\widehat{S}\left( \beta \right)$ as:

\begin{equation}
\widehat{S}\left( \tau ,\tau'\right) \equiv T\left\{ \exp \left[
-\int_{\tau'}^{\tau}d\tau_1 \widehat{H}_{s}\left( \tau_1 \right) \right]
\right\} .  \label{definosa}
\end{equation}


Clearly, $\widehat{S}\left( \beta ,0\right) =\widehat{S}\left( \beta \right)$. We can show that the quantity $\widehat{S}\left( \tau,\tau'\right)$ satisfies

\begin{equation}
\widehat{S}\left( \tau ,\tau ^{\prime }\right) =\widehat{1}-\int_{\tau
^{\prime }}^{\tau }d\tau _{1}\widehat{H}_{s}\left( \tau _{1}\right) \widehat{%
S}\left( \tau _{1},\tau ^{\prime }\right) .  \label{magica}
\end{equation}

Now, we functionally derive this expression with respect to the source $s_{j}\left( \mathbf{x},\tau _{x}\right) $:

\begin{align}
\frac{\delta \widehat{S}\left( \tau ,\tau ^{\prime }\right) }{\delta
s_{j}\left( \mathbf{x},\tau _{x}\right) } =&\,\theta \left( \tau -\tau _{x}\right) \theta \left( \tau _{x}-\tau
^{\prime }\right) \widehat{\phi }_{j}\left( \mathbf{x},\tau _{x}\right)
\widehat{S}\left( \tau _{x},\tau ^{\prime }\right) -\int_{\tau ^{\prime
}}^{\tau }d\tau _{1}\widehat{H}_{s}\left( \tau _{1}\right) \frac{\delta
\widehat{S}\left( \tau _{1},\tau ^{\prime }\right) }{\delta s_{j}\left(
\mathbf{x},\tau _{x}\right) }.
\end{align}

Choosing $\tau'=\tau_x$ in (\ref{magica}) and multiplying it by $\widehat{\phi }_{j}\left(
\mathbf{x},\tau _{x}\right) \widehat{S}\left( \tau _{x},\tau ^{\prime
}\right) $ results in

\begin{align}
\widehat{\phi }_{j}\left( \mathbf{x},\tau _{x}\right) \widehat{S}\left( \tau
_{x},\tau ^{\prime }\right) =&\,\widehat{S}\left( \tau ,\tau _{x}\right)
\widehat{\phi }_{j}\left( \mathbf{x},\tau _{x}\right) \widehat{S}\left( \tau
_{x},\tau ^{\prime }\right) +\int_{\tau _{x}}^{\tau }d\tau _{1}\widehat{H}%
_{s}\left( \tau _{1}\right) \widehat{S}\left( \tau _{1},\tau _{x}\right)\widehat{\phi }_{j}\left( \mathbf{x},\tau _{x}\right) \widehat{S}\left( \tau
_{x},\tau ^{\prime }\right) .
\end{align}

Therefore

\begin{align}
\frac{\delta \widehat{S}\left( \tau ,\tau ^{\prime }\right) }{\delta
s_{j}\left( \mathbf{x},\tau _{x}\right) }=&\,\left[\theta \left( \tau -\tau _{x}\right) \theta \left( \tau _{x}-\tau
^{\prime }\right) \widehat{S}\left( \tau ,\tau _{x}\right) \widehat{\phi }%
_{j}\left( \mathbf{x},\tau _{x}\right)\right.\notag\\
&\, +\theta \left( \tau -\tau _{x}\right) \theta \left( \tau _{x}-\tau
^{\prime }\right) \int_{\tau _{x}}^{\tau }d\tau _{1}\widehat{H}_{s}\left(
\tau _{1}\right)\left. \widehat{S}\left( \tau _{1},\tau _{x}\right) \widehat{\phi }%
_{j}\left( \mathbf{x},\tau _{x}\right) \right]\widehat{S}\left( \tau _{x},\tau
^{\prime }\right) \notag\\
&\,-\int_{\tau ^{\prime }}^{\tau }d\tau _{1}\widehat{H}_{s}\left( \tau
_{1}\right) \frac{\delta \widehat{S}\left( \tau _{1},\tau ^{\prime }\right)
}{\delta s_{j}\left( \mathbf{x},\tau _{x}\right) }.
\end{align}

The solution for this equation is

\begin{align}
\frac{\delta \widehat{S}\left( \tau ,\tau ^{\prime }\right) }{\delta
s_{j}\left( \mathbf{x},\tau _{x}\right) }=&\,\theta \left( \tau -\tau
_{x}\right) \theta \left( \tau _{x}-\tau ^{\prime }\right) \widehat{S}\left(
\tau ,\tau _{x}\right) \widehat{\phi }_{j}\left( \mathbf{x},\tau _{x}\right)
\widehat{S}\left( \tau _{x},\tau ^{\prime }\right) .
\end{align}

For any operator $\widehat{F}$, we define its dependence on the temperature in the presence of the external sources as
\begin{equation}
\widehat{F}^{s}\left( \tau \right) \equiv \widehat{\rho }_{s}^{-1}\left(
\tau \right) \,\widehat{F}\,\widehat{\rho }_{s}\left( \tau \right) .\label{icnostatus}
\end{equation}

This definition is very similar to the time evolution of an operator in the Heisenberg picture, with the density matrix taking the place of the time evolution operator. For later uses, we also compute the derivative of this definition with respect to $\tau$:\footnote{$\left[\widehat{A},\widehat{B}\right]\equiv\widehat{A}\widehat{B}-\widehat{B}\widehat{A}$.}

\begin{align}
\frac{\partial\widehat{F}^{s}\left( \tau \right)}{\partial\tau}=\left[\widehat{H}_T-\mu_j\widehat{N}_j,\widehat{F}^{s}\left( \tau \right)\right].\label{aparitus}
\end{align}

From the very definition of $\widehat{S}\left( \tau ,\tau'\right)$:

\begin{align}
\widehat{S}\left( \tau ,\tau_x\right)\widehat{S}\left( \tau_x ,0\right)=\widehat{S}\left( \tau ,0\right).
\end{align}

Then:

\begin{equation}
\frac{\delta \widehat{S}\left( \tau ,0\right) }{\delta s_{j}\left( \mathbf{x}%
,\tau _{x}\right) }=\theta \left( \tau -\tau _{x}\right) \widehat{S}\left(
\tau ,0\right) \widehat{\phi }_{j}^{s}\left( \mathbf{x},\tau _{x}\right) .\label{exquisite}
\end{equation}

From equations (\ref{rho em termos de rho zero}) and (\ref{exquisite}) with $0<\tau_x<\beta$, we get

\begin{equation}
\frac{\delta \widehat{\rho }_{s}\left( \beta \right) }{\delta s_{j}\left(
\mathbf{x},\tau _{x}\right) }=\widehat{\rho }_{s}\left( \beta \right)
\widehat{\phi }_{j}^{s}\left( \mathbf{x},\tau _{x}\right) .  \label{lama}
\end{equation}

Now, for an arbitrary operator $\widehat{F}^{s}\left( \tau \right)$ it is possible to show:

\begin{equation}
\frac{\delta }{\delta s_{j}\left( \mathbf{x},\tau _{x}\right) }\left[
\widehat{\rho }_{s}\left( \beta \right) \widehat{F}^{s}\left( \tau \right) %
\right] =\widehat{\rho }_{s}\left( \beta \right) T\left[ \widehat{\phi }%
_{j}^{s}\left( \mathbf{x},\tau _{x}\right) \widehat{F}^{s}\left( \tau
\right) \right] .  \label{splitizer}
\end{equation}

Hence, we find the result

\begin{equation}
\frac{\delta ^{2}\widehat{\rho }_{s}\left( \beta \right) }{\delta
s_{l}\left( \mathbf{y},\tau _{y}\right) \delta s_{j}\left( \mathbf{x},\tau
_{x}\right) } =\widehat{\rho }_{s}\left( \beta \right) T\left[ \widehat{\phi }%
_{l}^{s}\left( \mathbf{y},\tau _{y}\right) \widehat{\phi }_{j}^{s}\left(
\mathbf{x},\tau _{x}\right) \right] .  \label{barrenta}
\end{equation}

Let us define, now, the thermodynamical generating functional:

\begin{equation}
Z_{GF}\left[\left\{s_k\right\}\right]\equiv\mbox{Tr}\left[\widehat{\rho }_{s}\left( \beta \right)\right].\label{generating functional}
\end{equation}

From equation (\ref{partition}):

\begin{equation}
Z_{GF}\left[\left\{s_k=0\right\}\right]=Z\left(\beta\right).
\end{equation}

By taking the traces of equations (\ref{lama}) and (\ref{barrenta}), evaluating the results for the special case of vanishing sources, dividing them by the partition function and using (\ref{average}) and (\ref{generating functional}) we find:

\begin{align}
\left\langle\widehat{\phi }_{j}\left( \mathbf{x},\tau _{x}\right)\right\rangle=&\,\frac{1}{Z\left(\beta\right)} \left.\frac{\delta Z_{GF}\left[\left\{s_k\right\}\right] }{\delta s_{j}\left(
\mathbf{x},\tau _{x}\right) }\right|_{s=0};\label{leslie}\\
\left\langle \hspace{-.1cm}T\hspace{-.1cm}\left[ \widehat{\phi }%
_{l}\left( \mathbf{y},\tau _{y}\right)\hspace{-.05cm} \widehat{\phi }_{j}\left(
\mathbf{x},\tau _{x}\right)\hspace{-.05cm} \right]\hspace{-.05cm}\right\rangle\hspace{-.1cm}=&\,\hspace{-.05cm}\frac{1}{Z\left(\beta\right)} \hspace{-.1cm}\left.\frac{\delta ^{2}Z_{GF}\left[\left\{s_k\right\}\right] }{\delta s_{l}\hspace{-.1cm}\left( \mathbf{y},\tau _{y}\right) \delta s_{j}\hspace{-.1cm}\left( \mathbf{x},\tau_{x}\right) }\right|_{s=0}\hspace{-.05cm}.\label{nielsen}
\end{align}

From these results we see that we can compute thermal averages from the thermodynamical generating functional.

Before ending this section, we call attention to the fact that the $\tau$ variable in (\ref{leslie}) and (\ref{nielsen}) are restricted to the interval $\left(0,\beta\right)$. Outside this interval, the functional derivatives can be defined in several manners \cite{Fradkin1}. In section (\ref{DSF}) we will define the derivatives outside this interval in a consistent way.

\section{GQED$_4$ in thermodynamic equilibrium}\label{GQED}

In this section we consider the GQED$_4$ in thermal equilibrium. We restrict ourselves to the $\left(3+1\right)$ - dimensional case.

The Podolsky generalized classical electrodynamics is described by the following Lagrangian density:

\begin{equation}
\mathcal{L}_{GED}=\mathcal{L}_{D}+\mathcal{L}_{MC}+\mathcal{L}_P,\label{GED}
\end{equation}
where

\begin{align}
\mathcal{L}_{D} =&\,i\left( \gamma ^{\mu }\right) _{ab}\overline{\psi }_{a}\partial _{\mu }\psi
_{b} -m_{f}\overline{\psi }_{a}\psi _{a};\label{dirac}\\
 \mathcal{L}_{MC}=&\,-q_{e}\mathcal{A}_{\mu }\left(
\gamma ^{\mu }\right) _{ab}\overline{\psi }_{a}\psi _{b};\label{MC}\\
\mathcal{L}_P=&\,-\frac{1}{4}\mathcal{F}^{\mu\nu}\mathcal{F}_{\mu\nu} +\frac{1}{2m_P^2}\partial_\mu\mathcal{F}^{\mu\nu}\partial_\xi\mathcal{F}^{\xi}_{\phantom{\xi}\nu}.\label{podolsky}
\end{align}

Here, $ \gamma ^{\mu }$'s are the Dirac matrices satisfying

\begin{equation}
\left\{\gamma ^{\mu },\gamma ^{\nu }\right\}_{ab}=2g^{\mu\nu}\delta_{ab},
\end{equation}
$\psi_a=\psi_a(x)$ and $\overline{\psi }_{a}=\overline{\psi }_{a}(x)$ are Grassmannian fields, $a$ and $b$ run from 1 to 4, $m_{f}$ is an arbitrary parameter with dimension of energy (the so-called ``bare fermion mass"), $q_{e}$ is an arbitrary dimensionless parameter (the ``bare electrical charge") that plays the role the coupling constant of the theory, $\mathcal{A}_{\mu }=\mathcal{A}_{\mu }(x)$ is the Podolsky gauge field, $\mathcal{F}_{\mu\nu}\equiv \partial _{\mu }\mathcal{A}_{\nu }-\partial_{\nu }\mathcal{A}_{\mu }$, and $m_P$ is the Podolsky parameter, which is real and it has dimension of energy.\footnote{The parameter $m_P$ can be interpreted as the mass of the massive \textit{sector} of Podolsky field \cite{Bonin}.}

We notice that apart from the last term in the right-hand-side of (\ref{podolsky}), $\mathcal{L}_{GED}$ is the usual Lagrangian density for electrodynamics. The last term in (\ref{podolsky}) is a contribution from second-order derivatives of the electromagnetic field $\mathcal{A}_{\mu }$.

\subsection{The fermion fields}

As a first step towards a thermodynamical quantum theory, for now we consider only $\mathcal{L}_{D}+\mathcal{L}_{MC}$. This term is exact the same for both Maxwell and Podolsky theories. As it is well-known from the Dirac's constraints analysis of Maxwell electrodynamics, the only non-vanishing fundamental Dirac brackets between the fermionic fields only is \cite{Sundermeyer}

\begin{equation}
\left\{ \psi _{a}\left( x\right) ,\overline{\psi }_{b}\left(
y\right) \right\} _{D}^{x^0=y^0} =i\left( \gamma ^{0}\right) _{ab}\delta
\left( \mathbf{x}-\mathbf{y}\right).
\end{equation}

Replacing the Grassmannian functions by Grassmannian operators and using the correspondence principle we get:\footnote{$\left\{\widehat{A},\widehat{B}\right\}\equiv\widehat{A}\widehat{B}+\widehat{B}\widehat{A} $.}

\begin{equation}
\left\{ \widehat{\psi }_{a}\left( x\right) ,\widehat{\overline{\psi
}}_{b}\left( y\right) \right\}^{x^0=y^0} =-\left( \gamma ^{0}\right)
_{ab}\delta \left( \mathbf{x}-\mathbf{y}\right) \widehat{1}.\label{fun}
\end{equation}

As we have stated previously, stationarity is a requirement for thermodynamic equilibrium. This means no quantity can depend on time for a system in equilibrium. Since we are dealing with fields in thermodynamic equilibrium, we may evaluate  (\ref{fun}), say, for $x^0=y^0=0$ and simply rewrite it in a time-independent fashion:

\begin{equation}
\left\{ \widehat{\psi }_{a}\left( \mathbf{x}\right) ,\widehat{\overline{\psi
}}_{b}\left( \mathbf{y}\right) \right\} =-\left( \gamma ^{0}\right)
_{ab}\delta \left( \mathbf{x}-\mathbf{y}\right) \widehat{1}.\label{operator}
\end{equation}

The grand-canonical density matrix of the GQED$_4$ with external sources is

\begin{equation}
\widehat{\rho }_{s}\left( \beta \right) =\exp \left[ -\beta \left( \widehat{H}_{T}-\mu _{e}\widehat{N}\right) \right],\label{i cant take my mind off you}
\end{equation}
where

\begin{align}
\widehat{H}_{T} =&\,\widehat{H}_P+\widehat{H}_{D}+\widehat{H}_{MC}+\widehat{H}_{s};\\
\widehat{H}_{D} =&\,-\frac{1}{2}\int d^{3}z\left\{ i\left( \gamma ^{j}\right) _{ab}
\left[\widehat{\overline{\psi }}_{a}\left( \mathbf{z}\right) ,\partial _{j}^{\left(
z\right) }\widehat{\psi }_{b}\left( \mathbf{z}\right)
\right] -m_{f}\left[ \widehat{\overline{\psi }}_{a}\left(
\mathbf{z}\right), \widehat{\psi }_{a}\left( \mathbf{z}\right)\right]\right\};\\
\widehat{H}_{MC}=&\, \frac{q_{e}}{2}\left( \gamma ^{\mu }\right) _{ab}\int d^{3}z\widehat{\mathcal{A%
}}_{\mu }\left( \mathbf{z}\right) \left[ \widehat{\overline{\psi }}_{a}\left(
 \mathbf{z}\right), \widehat{\psi }_{b}\left(  \mathbf{z}\right)
\right] ; \\
\widehat{H}_{s}=&\,-\int d^{3}z\left\{ \mathcal{J}^{\mu }\left( \mathbf{z}\right) \widehat{%
\mathcal{A}}_{\mu }\left( \mathbf{z}\right) +\frac{1}{2}\left[\overline{\eta }_{a}\left(
\mathbf{z}\right) ,\widehat{\psi }_{a}\left( \mathbf{z}\right)\right]  +\frac{1}{2}\left[\eta
_{a}\left( \mathbf{z}\right), \widehat{\overline{\psi }}_{a}\left( \mathbf{z}\right)\right] \right\}.
\end{align}

$\widehat{H}_P$ is a Hamiltonian that commutes with the fermionic fields. $\mathcal{J}$, $\overline{\eta}$, and $\eta$ are the external sources. $\mu _{e}$ in (\ref{i cant take my mind off you}) is the chemical potential associated with the conserved charge $\widehat{N}$, which reads:

\begin{equation}
\widehat{N} =\frac{1}{2} \left( \gamma ^{0}\right) _{ab}\int d^{3}z\left[ \widehat{\overline{\psi }}%
_{a}\left( \mathbf{z}\right),\widehat{\psi }_{b}\left( \mathbf{z}\right)\right] .
\end{equation}



Now, we apply a similarity transformation with the density matrix to (\ref{operator}) and we use (\ref{icnostatus}) to write:

\begin{equation}
\left\{ \widehat{\psi }_{a}^s\left( \mathbf{x},\tau\right) ,\widehat{\overline{\psi
}}_{b}^s\left( \mathbf{y},\tau\right) \right\} =-\left( \gamma ^{0}\right)
_{ab}\delta \left( \mathbf{x}-\mathbf{y}\right) \widehat{1}.
\end{equation}

This is the fundamental anti-commutation relation for the fermionic part of (Generalized) QED in thermodynamic equilibrium. From this and from (\ref{aparitus}) we get the fermionic field equations:\footnote{From now on, implicit sums on greek indices are taken over an Euclidean geometry.}

\begin{align}
\left\{ \left( \gamma _{\mu }^{E}\right) _{ab}\widehat{D}_\mu^{\left(\mu_e,q_e\right)}\left[\widehat{A}^s\right]-m_{f}\delta _{ab}\right\} \widehat{\psi }_{b}^{s}\left( \mathbf{x},\tau \right) =&\,\eta _{a}\left( \mathbf{x},\tau \right) \widehat{1};\label{field equation psi}\\
\left\{ \left( \gamma _{\mu }^{E}\right) _{ba}\widehat{D}_\mu^{\left(-\mu_e,-q_e\right)}\left[\widehat{A}^s\right]+m_{f}\delta _{ba}\right\} \widehat{\overline{\psi }}%
_{b}^{s}\left( \mathbf{x},\tau \right) =&\,\overline{\eta }_{a}\left( \mathbf{x},\tau \right) \widehat{1},\label{field equation psi bar}
\end{align}
where we have defined the Euclidean Dirac matrices satisfying

\begin{equation}
\left\{ \gamma _{\mu }^{E},\gamma _{\nu }^{E}\right\}_{ab} =2\delta _{\mu \nu }\delta_{ab},\label{euclideandelta}
\end{equation}
and

\begin{align}
\widehat{D}_\mu^{\left(\mu_e,q_e\right)}\left[\widehat{A}^s\right] \equiv &\,\widehat{1}\partial_\mu^{\left(\mu_e\right)}+iq_e\widehat{A}_{\mu }^{s};\\
\partial_\mu^{\left(\mu_e\right)}\equiv&\,\partial_\mu+\mu_e\delta_{\mu 0};\\
\widehat{A}_{0}  \equiv &\,i\widehat{\mathcal{A}}^{0}; \label{euclidean0}\\
\widehat{A}_{j}  \equiv &\,-\widehat{\mathcal{A}}^{j} .\label{euclideanj}
\end{align}

First of all, we notice that the Euclidean spacetime structure of the quantum theory in thermodynamic equilibrium has emerged naturally in this formalism. Besides, we see that the field equation for $\widehat{\overline{\psi}}$ has opposite mass, charge and chemical potential signs if compared with the field equation for its conjugated field $\widehat{{\psi}}$.

\subsection{The Podolsky field}

Now, we apply the Nakanishi's auxiliary field method for the quantization of the Podolsky electromagnetic field. In order to do so, we must write a quantum Lagrangian density for the Podolsky field which breaks the gauge symmetry explicitly while preserving the Lorentz (or, for Euclidean field theories, the $SO\left(4\right)$) symmetry. For the GQED$_4$ in equilibrium, such a Lagrangian density is

\begin{align}
\widehat{\mathcal{L}}_{N} =\,&\frac{1}{4}\widehat{F}^s_{\mu \nu }\widehat{F}^s%
_{\mu \nu }\hspace{-.1cm}+\frac{1}{2m_{P}^{2}}\partial _{\mu }\widehat{F}^s_{\mu \nu
}\partial _{\xi }\widehat{F}^s_{\xi \nu }+\frac{1}{2}\left\{\widehat{B},G\left[\widehat{A}^s\right]\right\}-\frac{\alpha}{2} \widehat{B}^{2}+i\frac{q_{e}}{2}\widehat{A}^s_{\mu
}\left( \gamma _{\mu }^{E}\right) _{ab}\left[ \widehat{\overline{\psi }}^s_{a},\widehat{\psi }^s_{b}\right]+J_{\mu }\widehat{A}^s_{\mu },\label{L nakanishi}
\end{align}
where $\widehat{F}_{\mu \nu }\equiv\partial_\mu\widehat{A}_\nu-\partial_\nu\widehat{A}_\mu$, $\alpha$ is a non-vanishing real parameter (the covariant gauge parameter), $\widehat{B}$ is the auxiliary field, $G\left[\widehat{A}\right]$ is the gauge choice operator, and $J_\mu$ is the classical source.\footnote{From the Euclideanization we have $J_0=-i\mathcal{J}^0$ and $J_k=\mathcal{J}^k$.} By definition, under a gauge transformation, namely,

\begin{equation}
\widehat{A}_\mu\rightarrow\widehat{A}'_\mu=\widehat{A}_\mu+\partial_\mu\widehat{f}\label{gauge transformation}
\end{equation}
for any well-behaved operator $\widehat{f}$, the gauge choice operator must satisfy:

\begin{equation}
G\left[\widehat{A}\right]\rightarrow G\left[\widehat{A}'\right]\neq G\left[\widehat{A}\right].
\end{equation}

Due to this property, the quantum Lagrangian density $\widehat{\mathcal{L}}_{N}$ is not gauge invariant even if $q_e=0$ and $J_\mu=0$.

The thermodynamical quantum action associated with (\ref{L nakanishi}) is

\begin{equation}
\widehat{S}_N=\int_\beta d^4x\, \widehat{\mathcal{L}}_{N},\label{action}
\end{equation}
where we have used the notation

\begin{equation}
\int_\beta d^4x\equiv\int_0^\beta d\tau_x \int_V d^3x.
\end{equation}

The Schwinger principle for quantum theories states that under an arbitrary infinitesimal variation of the field operators the action (\ref{action}) behaves like

\begin{align}
\widehat{S}_N\rightarrow\widehat{S}_N'=\widehat{S}_N+\delta\widehat{S}_N
\end{align}
with
\begin{align}
\delta\widehat{S}_N=\int_\beta d^4x\partial_\mu \widehat{V}_\mu,
\end{align}
where $\widehat{V}_\mu$ is a vectorial operator that depends on both the fields and the variations \cite{Weiss}. Invoking this principle we find the field equations associated with the Lagrangian density (\ref{L nakanishi}):

\begin{align}
\left( \frac{\Delta }{m_{P}^{2}}+1\right) \partial _{\mu }\widehat{F}^s_{\mu
\nu }-\frac{\delta G^{\ast }\left[ \widehat{A}^s\right] }{\delta \widehat{A}^s_{\nu }}\widehat{B}=&\,i\frac{q_{e}}{2}\left( \gamma _{\nu }^{E}\right)
_{ab}\left[ \widehat{\overline{\psi }}^s_{a},\widehat{\psi }^s_{b}\right] +J_{\nu }\widehat{1};\label{limao}\\
\widehat{B}=&\,\frac{1}{\alpha }G\left[ \widehat{A}^s\right],\label{mamute}
\end{align}
where $\Delta\equiv -\partial_\mu\partial_\mu$. Here we have also defined

\begin{equation}
\widehat{B}\frac{\delta G\left[ \widehat{A}\right] }{ \delta \widehat{A}_{\mu }}\widehat{h}_{\mu}\equiv \frac{\delta G^{\ast }\left[ \widehat{A}\right] }{\delta \widehat{A}_{\mu} }\widehat{B}\,\widehat{h}_{\mu }+\partial _{\mu }\widehat{g}_{\mu },\label{palanque}
\end{equation}
where $\widehat{h}_{\mu}$ is any vectorial operator and $\widehat{g}_{\mu }$ is a suitable vectorial functional of the operators $\widehat{B}$, $ \widehat{A}_\mu$, and $\widehat{h}_{\mu }$.

The field equation for the auxiliary field (\ref{mamute}) is exactly solvable. By replacing the solution for $\widehat{B}$ in (\ref{limao}), its left-hand-side becomes

\begin{widetext}
\begin{align}
\left( \frac{\Delta }{m_{P}^{2}}+1\right) \partial _{\mu }\widehat{F}^s_{\mu\nu }-\frac{\delta G^{\ast }\left[ \widehat{A}^s\right] }{\delta \widehat{A}^s_{\nu }}\widehat{B} =-\left( \frac{\Delta
}{m_{P}^{2}}+1\right) \left( \delta _{\mu \nu }\Delta +\partial _{\nu
}\partial _{\mu }\right) \widehat{A}^s_{\mu }-\frac{1}{\alpha }\frac{\delta
G^{\ast }\left[ \widehat{A}^s\right] }{\delta \widehat{A}^s_{\nu }}G\left[
\widehat{A}^s\right].\label{my hips dont lie}
\end{align}
\end{widetext}

Just like in the Maxwellian case, the differential operator $\delta _{\mu \nu }\Delta +\partial _{\mu }\partial _{\nu}$ is not invertible. The next step in Nakanishi's procedure is finding an appropriate gauge choice $G\left[ \widehat{A}\right]$ in order to write the right-hand-side of (\ref{my hips dont lie}) as an invertible, local, differential operator acting on the gauge field. Inspired by the result found in the classical regime by one of us with a collaborator \cite{Pimentel}, we use a gauge choice based on the so-called \textit{generalized Lorenz condition}:

\begin{equation}
G\left[ \widehat{A}\right]=G_{GL}\left[ \widehat{A}\right]\equiv\left( \frac{\Delta }{m_P^2}+1\right) \partial_\mu \widehat{A}_\mu.\label{lorenz}
\end{equation}

With this choice, equation (\ref{palanque}) becomes

\begin{align}
\widehat{B}\left( \frac{\Delta }{m_P^2}+1\right) \partial_\mu \widehat{h}_{\mu}=&\, -\left( \frac{\Delta }{m_P^2}+1\right) \partial_\mu \widehat{B}\,\widehat{h}_{\mu }+\partial _{\mu }\widehat{g}_\mu,
\end{align}
where $\widehat{g}_\mu$ is unimportant for our purposes. This leads us to the identification

\begin{align}
\frac{\delta G^{\ast }\left[ \widehat{A}\right] }{\delta \widehat{A}_{\mu} }=-\left( \frac{\Delta }{m_P^2}+1\right) \partial_\mu .
\end{align}

With these results, we find the field equation for the Podolsky field:

\begin{align}
P_{\mu \nu }^{\left( m_{P}^{2},\alpha \right) } \widehat{A}^s_{\nu }\left( \mathbf{x},\tau \right) =&\, i\frac{q_{e}}{2}\left(\gamma_{\mu}^{E}\right)_{ab} \left[\widehat{\overline{\psi}}_{a}^s\left(\mathbf{x},\tau\right),\widehat{\psi}^s_{b}\left(\mathbf{x},\tau\right)\right]+J_{\mu}\left(\mathbf{x},\tau\right) \widehat{1},\label{field equation gauge}
\end{align}
where

\begin{widetext}
\begin{align}
P_{\mu \nu }^{\left( m_{P}^{2},\alpha \right) } \equiv-\left( \frac{\Delta }{m_{P}^{2}}+1\right) \left\{ \Delta \delta _{\mu \nu }+\left[ 1-\frac{1}{\alpha }\left( \frac{\Delta }{m_{P}^{2}}+1\right) \right]
\partial _{\mu }\partial _{\nu }\right\} .\label{Podolsky differential}
\end{align}
\end{widetext}

This differential operator differs greatly from its Maxwellian counterpart. For instance, in the Maxwell case, the Feynmann-St\"uckelberg gauge $\alpha=1$ makes the operator independent of the term $\partial_\mu\partial_\nu$. In the Podolsky case, there is no gauge in which that happens.

\subsection{The ghost fields}

As we have seen, the quantum Lagrangian density (\ref{L nakanishi}) with the gauge choice (\ref{lorenz}) is not invariant under an arbitrary gauge transformation (\ref{gauge transformation}). However, $\widehat{\mathcal{L}}_N$ with $q_e=0$ and $J_\mu=0$ is invariant under the following special gauge transformation

\begin{align}
\widehat{A}_{\mu } \rightarrow &\,\widehat{A}_{\mu }^{\prime }=\widehat{A}_{\mu }+\partial _{\mu }\widehat{\Lambda };  \\
\widehat{B} \rightarrow &\,\widehat{B}^{\prime }=\widehat{B},
\end{align}
provided the operator $\widehat{\Lambda }$ satisfies the following constraint

\begin{equation}
\left(\Delta+m_P^2\right)\Delta\widehat{\Lambda }=\widehat{0 }.\label{constraint}
\end{equation}

This is the \textit{residual gauge symmetry} of $\widehat{\mathcal{L}}_N$. It is not difficult to convince yourself that this residual gauge symmetry persists in the case $q_e\neq 0$ and $J_\mu$, provided we add the term corresponding to free fermions. This is precisely the case of our physical ensemble.

Now we implement the constraint (\ref{constraint}) in the theory by replacing (\ref{L nakanishi}) by

\begin{equation}
\widehat{\mathcal{L}}_C\equiv\widehat{\mathcal{L}}_N+\kappa \widehat{\lambda }\left( \mathbf{x},\tau \right) \left( \frac{\Delta }{m_{P}^{2}}+1\right) \Delta \widehat{\Lambda }\left( \mathbf{x},\tau \right).
\end{equation}

Here, $\widehat{\lambda }\left( \mathbf{x},\tau \right)$ is a Lagrange multiplier operator and $\kappa$ is a constant parameter whose value shall be fixed \textit{a fortiori}. Now we rewrite the Lagrange multiplier as

\begin{equation}
\widehat{\lambda }\left( \mathbf{x},\tau \right) \equiv i\,\widehat{\overline{C}}\left( \mathbf{x},\tau \right) \upsilon.
\end{equation}
where $\widehat{\overline{C}}\left( \mathbf{x},\tau \right)$ is a Grassmannian field operator and $\upsilon$ is a Grassmannian constant. So, up to a total derivative of a vectorial operator, we can write

\begin{equation}
\widehat{\mathcal{L}}_{C}=\widehat{\mathcal{L}}_{N}+i\kappa \partial _{\mu } \widehat{%
\overline{C}}\left( \mathbf{x},\tau \right) \left( \frac{%
\overleftarrow{\partial }_{\mu }\overrightarrow{\partial }_{\nu }}{m_{P}^{2}}+\delta _{\mu \nu }\right) \partial _{\nu }\widehat{C}\left( \mathbf{x},\tau \right) ,\label{indident}
\end{equation}
where $\widehat{C}\left( \mathbf{x},\tau \right)\equiv\upsilon\widehat{\Lambda}\left( \mathbf{x},\tau \right)$ is another Grassmannian field operator. $\widehat{C}$ and $\widehat{\overline{C}}$ are called \textit{ghost fields} \cite{Nakanishi}.

In order to find the field equations for the ghost fields, we add the Grassmannian fields $d\left( \mathbf{x},\tau \right)$ and $\overline{d}\left( \mathbf{x},\tau \right)$ as sources for them:

\begin{equation}
\widehat{\mathcal{L}}_{gs}\equiv\widehat{\mathcal{L}}_{C}+\frac{1}{2}\left[\overline{d},\widehat{C}^s\right] +\frac{1}{2}\left[d,\widehat{\overline{C}}^s\right].
\end{equation}

From this quantum Lagrangian density, we find the ghost field equations:

\begin{align}
i\kappa \left( \frac{\Delta }{m_{P}^{2}}+1\right) \Delta \widehat{C}^s\left(
\mathbf{x},\tau \right) =&\,d\left( \mathbf{x},\tau \right) \widehat{1}; \label{caju}\\
i\kappa \left( \frac{\Delta }{m_{P}^{2}}+1\right) \Delta \widehat{\overline{C}}^s\left( \mathbf{x},\tau \right) =&\,-\overline{d}\left( \mathbf{x},\tau
\right) \widehat{1}.\label{jacu}
\end{align}

We have introduced the ghost fields to take the constraint (\ref{constraint}) from the residual gauge symmetry into account in the GQED$_4$. However, by introducing these news fields, we have incidentally introduced a new symmetry in the theory. The quantum Lagrangian density (\ref{indident}) is invariant under the following \textit{global} symmetry:

\begin{align}
\widehat{C}\left(\mathbf{x},\tau \right)\rightarrow&\,\widehat{C}'\left(\mathbf{x},\tau \right)=e^{i\theta_0}\widehat{C}\left(\mathbf{x},\tau\right);\\
\widehat{\overline{C}}\left(\mathbf{x},\tau \right)\rightarrow&\,\widehat{\overline{C}}'\left(\mathbf{x},\tau \right)=\widehat{\overline{C}}\left(\mathbf{x},\tau \right)e^{-i\theta_0},
\end{align}
where $\theta_0$ is a real, constant, arbitrary number. Associated with this continuous global symmetry, we have the following conserved charge:

\begin{align}
\widehat{Q} =&\,-\frac{i}{2}\int d^{3}z\left\{ \left[\widehat{\overline{\pi }},\widehat{C}\right] +\left[\widehat{\overline{C}},\widehat{\pi}\right] +\left[ \widehat{\overline{P}}, \widehat{D}\right] +\left[\widehat{\overline{D}}, \widehat{P}\right]\right\},
\end{align}
where $\widehat{\overline{\pi }}=\widehat{\overline{\pi }}\left(\mathbf{z},\tau\right)$ and $\widehat{{\pi}}=\widehat{{\pi}}\left(\mathbf{z},\tau\right)$ are the conjugated momentum operators to $\widehat{C}$ and $\widehat{\overline{C}}$  respectively and $\widehat{\overline{P }}=\widehat{\overline{P }}\left(\mathbf{z},\tau\right)$ and $\widehat{{P}}=\widehat{{P}}\left(\mathbf{z},\tau\right)$ are the conjugated momentum operators to $\widehat{D}=\widehat{D}\left(\mathbf{z},\tau\right)$ and $\widehat{\overline{D}}=\widehat{\overline{D}}\left(\mathbf{z},\tau\right)$. The extra fields $\widehat{D}$ and $\widehat{\overline{D}}$ appear from the Ostrogradsky formalism for theories of second-order derivatives and are defined in the classical level before the Euclideanization as  ${D}\equiv\partial{C}/\partial t$ and ${\overline{D}}\equiv\partial{\overline{C}}/\partial t$ \cite{Ostrogradsky}. They are treated as independent of the original ghost fields. The Noether operator $\widehat{Q}$ is called the \textit{ghost charge}.

Since we have a new Noether operator in the problem, we must take it into account in the density matrix. Therefore, instead of (\ref{i cant take my mind off you}), the true partition function is \cite{Hata Kugo}

\begin{equation}
\widehat{\rho }_{gs}\left( \beta \right) =\exp \left[ -\beta \left( \widehat{H}_{T}-\mu _{e}\widehat{N}-\mu_g\widehat{Q}\right) \right],\label{ictiossaurus}
\end{equation}
where the total Hamiltonian is modified in order to take into account the free Hamiltonian for the ghost fields and the source Hamiltonian for the ghost sources. In this expression, $\mu_g$ is the \textit{ghost chemical potential}, \textit{i.e.}, the chemical potential associated with the ghost charge operator. We can also define the dependence of an arbitrary operator $\widehat{F}$ with the temperature in the ensemble described by this new density matrix. It reads:

\begin{align}
\widehat{F}^{gs}\left(\beta\right)\equiv\widehat{\rho}_{gs}^{-1}\left(\beta\right)\widehat{F}\,\widehat{\rho}_{gs}\left(\beta\right).
\end{align}

However, for a field operator $\widehat{w}_j$ which commutes with the ghost fields, we have:

\begin{align}
\widehat{w}_j^{gs}\left(\beta\right)\hspace{-.1cm}=\widehat{\rho}_{gs}^{-1}\left(\beta\right)\widehat{w}_j\,\widehat{\rho}_{gs}\left(\beta\right)
=\widehat{\rho}_{s}^{-1}\left(\beta\right)\widehat{w}_j\,\widehat{\rho}_{s}\left(\beta\right)\hspace{-.1cm}=\widehat{w}_j^{s}\left(\beta\right),
\end{align}
with $\widehat{\rho}_{s}\left(\beta\right)$ given by (\ref{i cant take my mind off you}). This is the case of the fermionic fields $\widehat{\psi}_a$ and $\widehat{\overline{\psi}}_a$ and of the Podolsky field $\widehat{A}_\mu$ and it explains why we can use the density matrix (\ref{i cant take my mind off you}) for these fields. However, for the ghost fields themselves:

\begin{align}
\widehat{C}^{gs}\left( \mathbf{x},\tau \right) =&\,e^{\tau \mu _{g}}\widehat{C}^{s}\left( \mathbf{x},\tau \right);\label{para c}\\
\widehat{\overline{C}}^{gs}\left( \mathbf{x},\tau \right) =&\,e^{-\tau \mu
_{g}}\widehat{\overline{C}}^{s}\left( \mathbf{x},\tau \right).\label{para c barra}
\end{align}

Here, and in field equations (\ref{caju}) and (\ref{jacu}) as well, $\widehat{C}^{s}$ and $\widehat{\overline{C}}^{s}$ are obtained from a similarity transformation with the density matrix (\ref{ictiossaurus}) evaluated with $\mu_g=0$.

\section{The periodicity of some special thermal averages}\label{Green}
From the field equations (\ref{field equation psi}), (\ref{field equation psi bar}), (\ref{field equation gauge}), (\ref{caju}), and (\ref{jacu}) we can find a set of functional equations satisfied by the thermodynamical generating functional of the Podolsky theory for the electrodynamics. By multiplying each of these field equations by the density matrix (\ref{ictiossaurus}), making use of (\ref{lama}) and (\ref{barrenta}), taking the trace of the results and using the definition (\ref{generating functional}), we find:

\begin{widetext}
\begin{align}
\left[ \left( \gamma _{\mu }^{E}\right) _{ab}\partial _{\mu }^{\left( \mu
_{e}\right) }-m_{f}\delta _{ab}\right]_x \frac{\delta Z_{GF}}{\delta \overline{\eta }_{b}\left(
\mathbf{x},\tau_x \right) } =&\,\left[ -iq_{e}\left( \gamma _{\nu }^{E}\right)
_{ab}\frac{\delta ^{2}}{\delta J_{\nu }\left( \mathbf{x},\tau_x \right) \delta
\overline{\eta }_{b}\left( \mathbf{x},\tau_x \right) }+\eta _{a}\left( \mathbf{x},\tau_x \right) \right]  Z_{GF} ;\label{leite} \\
\left[ \left( \gamma _{\mu }^{E}\right) _{ba}\partial _{\mu }^{\left( -\mu
_{e}\right) }+m_{f}\delta _{ba}\right]_x \frac{\delta Z_{GF}}{\delta \eta _{b}\left( \mathbf{x}%
,\tau_x \right) } =&\,\left[ iq_{e}\left( \gamma _{\nu }^{E}\right) _{ba}\frac{%
\delta ^{2}}{\delta J_{\nu }\left( \mathbf{x},\tau_x \right) \delta \eta
_{b}\left( \mathbf{x},\tau_x \right) }+\overline{\eta }_{a}\left( \mathbf{x}%
,\tau_x \right) \right] Z_{GF} ;\label{lesma} \\
P_{\mu \nu }^{\left( m_{P}^{2},\alpha \right) }\left( x\right) \frac{\delta
Z_{GF} }{\delta J_{\nu}\left( \mathbf{x},\tau_x \right) } =&\,\left[ iq_{e}\left( \gamma _{\mu
}^{E}\right) _{ab}\frac{\delta ^{2}}{\delta \eta _{a}\left( \mathbf{x},\tau_x
\right) \delta \overline{\eta }_{b}\left( \mathbf{x},\tau_x \right) }+J_{\mu
}\left( \mathbf{x},\tau_x \right) \right]Z_{GF} ;\label{pastel} \\
i\kappa \left( \frac{\Delta }{m_{P}^{2}}+1\right) \Delta_x \frac{\delta Z_{GF} }{\delta \overline{d}%
\left( \mathbf{x},\tau_x \right) } =&\,d\left( \mathbf{x},\tau_x \right) Z_{GF} ;  \label{azulejo} \\
i\kappa \left( \frac{\Delta }{m_{P}^{2}}+1\right) \Delta_x \frac{\delta Z_{GF}}{\delta d\left(
\mathbf{x},\tau_x \right) } =&\,-\overline{d}\left( \mathbf{x},\tau_x \right)Z_{GF} ,\label{pedrita}
\end{align}
\end{widetext}
where $Z_{GF} =Z_{GF}\left[ J,\eta ,\overline{\eta },d,\overline{d}\right] $ is the thermodynamical generating functional. The whole physical content of the theory is encrypted in this set of coupled functional equations. It is possible to solve this set of equations, obtaining in this way the generating functional. However, for our purposes, we shall first rewrite this equations in a more convenient form. By properly deriving each of these equations once more, we can replace this set of coupled functional equations by a set of differential equations interconnecting several thermal averages of the fields:

\begin{widetext}
\begin{align}
\left[ \left( \gamma _{\mu }^{E}\right) _{ab}\partial _{\mu }^{\left( \mu
_{e}\right) }-m_{f}\delta _{ab}\right] _{x}\left\langle T\left[ \widehat{%
\overline{\psi }}_{c}\left( \mathbf{y},\tau _{y}\right) \widehat{\psi }%
_{b}\left( \mathbf{x},\tau _{x}\right) \right] \right\rangle
=&\,-iq_{e}\left( \gamma _{\nu }^{E}\right) _{ab}\notag\\
&\,\times\left\langle T\left[
\widehat{\overline{\psi }}_{c}\left( \mathbf{y},\tau _{y}\right) \widehat{A}%
_{\nu }\left( \mathbf{x},\tau _{x}\right) \widehat{\psi }_{b}\left( \mathbf{x%
},\tau _{x}\right) \right] \right\rangle \notag\\
&\,+\delta _{ac}\delta \left( \mathbf{x}-\mathbf{y}\right) \delta \left( \tau
_{x}-\tau _{y}\right) ;\label{paqui}\\
\left[ \left( \gamma _{\mu }^{E}\right) _{ba}\partial _{\mu }^{\left( -\mu
_{e}\right) }+m_{f}\delta _{ba}\right] _{x}\left\langle T\left[ \widehat{%
\psi }_{c}\left( \mathbf{y},\tau _{y}\right) \widehat{\overline{\psi }}%
_{b}\left( \mathbf{x},\tau _{x}\right) \right] \right\rangle
=&\,iq_{e}\left( \gamma _{\nu }^{E}\right) _{ba}\notag\\
&\,\times\left\langle T\left[ \widehat{%
\psi }_{c}\left( \mathbf{y},\tau _{y}\right) \widehat{A}_{\nu }\left(
\mathbf{x},\tau _{x}\right) \widehat{\overline{\psi }}_{b}\left( \mathbf{x}%
,\tau _{x}\right) \right] \right\rangle \notag\\
&\,+\delta _{ac}\delta \left( \mathbf{x}-\mathbf{y}\right) \delta \left( \tau
_{x}-\tau _{y}\right) ;  \\
P_{\mu \nu }^{\left( m_{P}^{2},\alpha \right) }\left( x\right) \left\langle T
\left[ \widehat{A}_{\xi }\left( \mathbf{y},\tau _{y}\right) \widehat{A}_{\nu
}\left( \mathbf{x},\tau _{x}\right) \right] \right\rangle
=&\,iq_{e}\left( \gamma _{\mu }^{E}\right) _{ab}\notag\\
&\,\times\left\langle T\left[ \widehat{%
A}_{\xi }\left( \mathbf{y},\tau _{y}\right) \widehat{\overline{\psi }}%
_{a}\left( \mathbf{x},\tau _{x}\right) \widehat{\psi }_{b}\left( \mathbf{x}%
,\tau _{x}\right) \right] \right\rangle \notag\\
&\,+\delta _{\mu \xi }\delta \left( \mathbf{x}-\mathbf{y}\right) \delta
\left( \tau _{x}-\tau _{y}\right) ;  \\
i\kappa \left( \frac{\Delta}{m_{P}^{2}}+1\right) \Delta
_{x}\left\langle T\left[ \widehat{\overline{C}}\left( \mathbf{y},\tau
_{y}\right) \widehat{C}\left( \mathbf{x},\tau _{x}\right) \right]
\right\rangle =&\,\delta \left( \mathbf{x}-\mathbf{y}\right) \delta
\left( \tau _{x}-\tau _{y}\right) ;  \label{paCao}\\
i\kappa \left( \frac{\Delta }{m_{P}^{2}}+1\right) \Delta
_{x}\left\langle T\left[ \widehat{C}\left( \mathbf{y},\tau _{y}\right)
\widehat{\overline{C}}\left( \mathbf{x},\tau _{x}\right) \right]
\right\rangle=&\,-\delta \left( \mathbf{x}-\mathbf{y}\right) \delta
\left( \tau _{x}-\tau _{y}\right) .  \label{paCola}
\end{align}
\end{widetext}

Let us consider ensemble averages of the form  $\left\langle T\left[ \widehat{\phi }_{r'}\left( \mathbf{y},\tau_{y}\right) \widehat{\phi}_{r}\left(\mathbf{x},\tau _{x}\right) \right]
\right\rangle$, where $\widehat{\phi }_r\left( \mathbf{x},\tau_{x}\right)$ is any of the fields. Using  the definitions of both the ordering of two operators (\ref{lamel}) and the thermal average (\ref{average}) we write

\begin{widetext}
\begin{align}
\left\langle T\left[ \widehat{\phi }_{r'}\left( \mathbf{y},\tau _{y}\right)
\widehat{\phi}_{r}\left( \mathbf{x},\tau _{x}\right) \right] \right\rangle=&\,\theta \left( \tau _{y}-\tau _{x}\right) \frac{\text{Tr}\left[ \widehat{\rho}_{g}\left( \beta \right) \widehat{\phi }_{r'}\left( \mathbf{y},\tau
_{y}\right) \widehat{\phi}_{r}\left( \mathbf{x},\tau _{x}\right) \right] }{Z\left(\beta \right) }  \notag\\
&\,\pm \theta \left( \tau _{x}-\tau _{y}\right) \frac{\text{Tr}\left[
\widehat{\rho }_{g}\left( \beta \right) \widehat{\phi}_{r}\left( \mathbf{x},\tau
_{x}\right) \widehat{\phi }_{r'}\left( \mathbf{y},\tau _{y}\right) \right] }{Z\left( \beta \right) },\label{lindinha}
\end{align}
\end{widetext}
where $\widehat{\rho }_{g}\left( \beta \right)$ is the partition function (\ref{ictiossaurus}) evaluated for vanishing classical sources.

Using an adaptation of equation (\ref{without}) and the fact that the momentum operator $ \widehat{\mathbf{P}}$ is the generator of the spatial translations, we can write:

\begin{align}
\widehat{\phi }_{r}\left( \mathbf{x},\tau _{x}\right) =&\,\widehat{\rho }%
_{g}^{-1}\left( \tau _{x}\right) \widehat{\phi }_{r}\left( \mathbf{x}%
,0\right) \widehat{\rho }_{g}\left( \tau _{x}\right) ;\\
\widehat{\phi }_{r}\left( \mathbf{x},0\right) =&\,e^{-i\mathbf{x}\cdot \widehat{%
\mathbf{P}}}\widehat{\phi }_{r}\left( \mathbf{0},0\right) e^{i\mathbf{x}%
\cdot \widehat{\mathbf{P}}}.
\end{align}

Now we can compute the traces of the right-hand-side of (\ref{lindinha}) in any basis. Without loss of generality, let us consider the mutual eigenbasis for energy, momentum and Noether charges $\left\vert E,\mathbf{P},N,N_g\right\rangle$. Using this basis for the traces computation, we find that (\ref{lindinha}) depends neither on the absolute values of the spatial coordinates $\mathbf{x}$ and $\mathbf{y}$ nor on the absolute values of the variables related with the temperature $\tau_x$ and $\tau_y$, but only on both the differences $\mathbf{x}-\mathbf{y}$ and $\tau_x-\tau_y$. It is, then, convenient to define the following quantities:

\begin{align}
\left\langle T\left[ \widehat{A}_{\nu }\left( \mathbf{y},\tau _{y}\right)
\widehat{A}_{\mu }\left( \mathbf{x},\tau _{x}\right) \right] \right\rangle
 \equiv &\,D_{\mu \nu }\left( \mathbf{x}-\mathbf{y},\tau _{x}-\tau
_{y}\right) ;  \label{limbo} \\
\left\langle T\left[ \widehat{\overline{\psi }}_{b}\left( \mathbf{y},\tau
_{y}\right) \widehat{\psi }_{a}\left( \mathbf{x},\tau _{x}\right) \right]
\right\rangle  \equiv &\,S_{ab}\left( \mathbf{x}-\mathbf{y},\tau _{x}-\tau
_{y}\right) ;\label{limboso} \\
\left\langle T\left[ \widehat{\overline{C}}\left( \mathbf{y},\tau
_{y}\right) \widehat{C}\left( \mathbf{x},\tau _{x}\right) \right]
\right\rangle  \equiv &\,G\left( \mathbf{x}-\mathbf{y},\tau _{x}-\tau
_{y}\right). \label{aluminio transparente}
\end{align}

From the definition of the ordering operator, we also find

\begin{eqnarray}
\left\langle T\left[ \widehat{A}_{\mu }\left( \mathbf{x},\tau _{x}\right)\widehat{A}_{\nu }\left( \mathbf{y},\tau _{y}\right) \right] \right\rangle
 &=&D_{ \nu \mu}\left( \mathbf{y-x}%
,\tau _{y}-\tau _{x}\right);   \\
\left\langle T\left[ \widehat{\psi }_{b}\left( \mathbf{y},\tau _{y}\right)
\widehat{\overline{\psi }}_{a}\left( \mathbf{x},\tau _{x}\right) \right]
\right\rangle&=&-S_{ba}\left( \mathbf{y-x}%
,\tau _{y}-\tau _{x}\right) ;   \\
\left\langle T\left[ \widehat{C}\left( \mathbf{y},\tau _{y}\right) \widehat{%
\overline{C}}\left( \mathbf{x},\tau _{x}\right) \right] \right\rangle&=&-G\left( \mathbf{y-x},\tau _{y}-\tau _{x}\right) .
\label{pessego}
\end{eqnarray}

Since the electromagnetic field commutes with the ghost fields, using (\ref{lindinha}) for $0<\tau<\beta$, we find

\begin{eqnarray}
D_{\mu \nu }\left( \mathbf{x}-\mathbf{y},\tau \right) &=&\left\langle T\left[ \widehat{A
}_{\nu}\left( \mathbf{y},0\right) \widehat{A}_{\mu}\left( \mathbf{x},\tau
\right) \right] \right\rangle   \notag \\
&=& \frac{\text{Tr}\left[ \widehat{\rho }_{g}\left( \beta \right)
\widehat{A }_{\mu}\left( \mathbf{x},\tau \right) \widehat{A}_{\nu}\left(
\mathbf{y},0\right) \right] }{Z\left( \beta \right) } \notag \\
&=& \frac{\text{Tr}\left[ \widehat{\rho }_{g}\left( \beta \right)
\widehat{A}_{\nu}\left( \mathbf{y},\beta \right) \widehat{A}_{\mu}\left(
\mathbf{x},\tau \right) \right] }{Z\left( \beta ,V,\mu _{e}\right) }\notag\\
&=& D_{\mu \nu }\left( \mathbf{x}-\mathbf{y},\tau -\beta \right) .\label{va}
\end{eqnarray}

Similarly, taking into account the Grassmannian nature of the fermionic fields and the fact that they also commute with the ghost fields, we have:

\begin{align}
S_{ab}\left( \mathbf{x}-\mathbf{y},\tau \right)=-S_{ab}\left( \mathbf{x}-\mathbf{y},\tau-\beta \right).\label{ve}
\end{align}

So, we see that the thermal averages of the ordering of fields imply the function $D_{\mu\nu}$ must be \textit{periodic} in the $\tau$ coordinate while the function $S_{ab}$ must be \textit{anti-periodic}.

However, repeating these steps for the quantity (\ref{aluminio transparente}) and using expressions (\ref{para c}) and (\ref{para c barra}) with vanishing sources, we find

\begin{align}
G\left( \mathbf{x}-\mathbf{y},\tau \right) =&\,-e^{-\beta \mu _{g}}G\left( \mathbf{x}-\mathbf{y},\tau -\beta \right) ;\label{bacteria}\\
G\left( \mathbf{y-x},-\tau \right) =&\,-e^{\beta \mu _{g}}G\left( \mathbf{y-x},\beta -\tau\right) .  \label{arquea}
\end{align}

Equations (\ref{paCao}) and (\ref{paCola}), (\ref{aluminio transparente}) and (\ref{pessego}) show us that both $G\left(\mathbf{x}-\mathbf{y},\tau _{x}-\tau _{y}\right)$ and $G\left(\mathbf{y-x},\tau _{y}-\tau _{x}\right) $ are the Green functions of the operator
$i\kappa \left( {\Delta }/m_{P}^{2}+1\right) \Delta$. This means these two quantities must be equal to each other. Hence, it follows:

\begin{equation}
G\left( \mathbf{x}-\mathbf{y},\tau -\beta\right) =e^{2\beta \mu_g}G\left( \mathbf{x}-\mathbf{y},\tau -\beta \right) .\label{gezao}
\end{equation}

Besides the trivial solution $\mu_g=0$, the above equation shows us that any ghost chemical potential satisfying

\begin{equation}
\mu_g=\frac{in\pi}{\beta},\label{muzinho}
\end{equation}
with integer $n$, is physically admissible. This means that a non-vanishing ghost chemical potential is not a real number and, therefore, it is not an observable thermodynamical quantity.

Equation (\ref{bacteria}) shows that the ghost Green function $G\left( \mathbf{x},\tau \right)$ can be either periodic or anti-periodic in the interval $\beta$, depending on whether we choose $n$ in the chemical potential (\ref{muzinho}) to be odd or even, respectively \cite{Hata Kugo}.

\section{The partition function}\label{sec partition function}

In the present section we will seek path integral representations for the thermodynamical generating functional and for the partition function of the theory. In order to achieve that goal, we try the following path integral representation for the generating functional:

\begin{widetext}
\begin{eqnarray}
Z_{GF}\left[ J,\eta ,\overline{\eta },d,\overline{d}\right] &=&\int_{P}%
\mathcal{D}A\int_{A-P}\mathcal{D}\overline{\psi }\mathcal{D}\psi \int_{\mu_g}%
\mathcal{D}\overline{C}\mathcal{D}C\,\widetilde{Z}_{GF}\left[ A,\psi ,%
\overline{\psi },C,\overline{C}\right]  \notag \\
&&\times \exp \left[ \int_{\beta }d^{4}x\left( J_{\mu }A_{\mu }+\overline{%
\eta }_{a}\psi _{a}-\overline{\psi }_{a}\eta _{a}-\overline{C}d+\overline{d}%
C\right) \right] ,\label{functional fourier}
\end{eqnarray}
\end{widetext}
where $\mathcal{D}A\equiv \prod\limits_{\sigma =0}^{3}%
\mathcal{D}A_{\sigma }$ and $\mathcal{D}\overline{\psi }\mathcal{D}\psi \equiv
\prod\limits_{a=1}^{4}\mathcal{D}\overline{\psi }_{a}\mathcal{D}\psi _{a}$. $P$ stands for integration over all periodic configurations of the Podolsky field $A_\mu\left(\mathbf{x},0\right)=A_\mu\left(\mathbf{x},\beta\right)$ and $A-P$ for all anti-periodic configurations $\psi_a\left(\mathbf{x},0\right)=-\psi_a\left(\mathbf{x},\beta\right)$ and $\overline{\psi}_a\left(\mathbf{x},0\right)=-\overline{\psi}_a\left(\mathbf{x},\beta\right)$. These periodicity conditions are chosen in order to satisfy the properties (\ref{va}) for the Podolsky field and (\ref{ve}) for the fermionic fields. The integration over the ghost fields $C$ and $\overline{C}$ depends on the choice for the ghost chemical potential (\ref{muzinho}) through  (\ref{bacteria}).

By replacing (\ref{functional fourier}) in any equation of the set of functional equations (\ref{leite}-\ref{pedrita}), we find the following

\begin{equation}
\frac{\delta \widetilde{Z}_{GF} }{\delta \phi_{j }\left( \mathbf{x},\tau \right) }=-\frac{\delta
S_T }{\delta \phi_{j }\left( \mathbf{x},\tau \right) }\widetilde{Z}_{GF} ,\label{solve}
\end{equation}
where $\phi_{j }$ stands for any of the fields and $S_T=S_T^{\left( \beta ,\mu _{e},V\right) }\left[ A,\psi ,\overline{\psi },C,\overline{C}\right]$ is the \textit{thermodynamical action}:

\begin{widetext}
\begin{align}
S_T\equiv&\,\int_\beta d^4x \left\{\frac{1}{2}A_\mu P_{\mu \nu }^{\left( m_{P}^{2},\alpha \right) } A_\nu
-i\kappa \partial _{\mu }\overline{C}\left( \delta _{\mu \nu }+\frac{\overleftarrow{\partial }_{\mu }\overrightarrow{\partial }_{\nu }}{m_{P}^{2}}%
\right) \partial _{\nu }C\right.\notag\\
&\,\left.-\overline{\psi }_{a}\left[ \left( \gamma _{\mu
}^{E}\right) _{ab}\partial _{\mu }^{\left( \mu _{e}\right) }-m_{f}\delta
_{ab}\right] \psi _{b} -iq_{e}A_{\mu }\overline{\psi }_{a}\left( \gamma _{\mu }^{E}\right)
_{ab}\psi _{b}\right\}.\label{thermodynamical action}
\end{align}
\end{widetext}

The solution for the functional equation (\ref{solve}) is

\begin{equation}
\widetilde{Z}_{GF}\left[ A,\psi ,\overline{\psi },C,\overline{C}\right] =\widetilde{Z}_{0}
e^{-S_T^{\left( \beta ,\mu _{e},V\right) }\left[ A,\psi ,\overline{\psi },C,\overline{C}\right] }.
\end{equation}

 $\widetilde{Z}_{0}$ is some so far unknown constant.

By setting the sources equal to zero in equation (\ref{functional fourier}), we formally get the partition function of the GQED$_4$. However, we still have to fix the value of the ghost chemical potential in order to perform the path integration over the ghost fields.

First of all, we notice that the periodicity conditions imposed over the functional integrals do not depend whether we are dealing with interacting or with free fields. Taking this property into account, we specialize ourselves in the free case, \textit{i.e.}, with $q_e=0$. It is possible to compute the free partition function exactly:

\begin{align}
Z_F\left(\beta\right)=&\,\widetilde{Z}_{0}\det_{P}\left[ \frac{1}{\alpha }\left( \frac{\Delta }{%
m_{P}^{2}}+1\right) \right] ^{-\frac{1}{2}}\hspace{-.2cm}\det_{P}\left[ \left( \frac{%
\Delta }{m_{P}^{2}}+1\right) \Delta \right] ^{-2}\det_{\mu_e}\left[ \kappa
\left( \frac{\Delta }{m_{P}^{2}}+1\right) \Delta \right]    \notag \\
&\,\times \det_{A-P}\left[ \left( \gamma _{\mu }^{E}\right) _{ab}\partial
_{\mu }^{\left( \mu _{e}\right) }-m_{f}\delta _{ab}\right] .
\end{align}

Here, $\det_{P}$ stands for the determinant evaluated over periodic funcions, $\det_{A-P}$ for the determinant evaluated over anti-periodic functions, and $\det_{\mu_e}$ for the determinant over functions that can be either periodic or anti-periodic depending on the choice of the ghost chemical potential. From this expression we see that the operator that the ghost determinant is evaluated upon appears also in the periodic determinant that comes from the Podolsky field, but does not appear in the fermionic, anti-periodic determinant. For this reason, we choose the ghost chemical potential (\ref{muzinho}) to be with an odd $n$ \cite{Hata Kugo}. If that is assumed, we can rewrite the free partition function as

\begin{align}
Z_F\left(\beta\right)=Z_P\left(\beta\right)Z_D\left(\beta\right),
\end{align}
where $Z_D\left(\beta\right)=\det_{A-P}\left[ \left( \gamma _{\mu }^{E}\right) _{ab}\partial _{\mu
}^{\left( \mu _{e}\right) }-m_{f}\delta _{ab}\right]$ is the partition function for free fermions \cite{Kapusta} and

\begin{align}
Z_P\left(\beta\right)=\det_{P}\left( \Delta
+m_{P}^{2}\right) ^{-\frac{3}{2}}\det_{P}\left( \Delta \right)^{-1}\label{economia}
\end{align}
is the partition function for the free Podolsky field, which describes a gas of non-interacting massless vectorial particles with Proca particles with mass $m_P$. This result coincides with the one found by us in our previous work with the imaginary-time method \cite{Bonin}. In writing equation (\ref{economia}) we have found the values of the two constants that were still unknown, namely, $\kappa=\alpha^{-1/2}$ and $\widetilde{Z}_{0}=\det_P\left(m_P\right)^{-3}$.

Using Nakanishi's method and the Matsubara-Fradkin formalism we have found the correct partition function for the free Podolsky field \textit{without} performing its  Dirac constraints analysis.

With these new results, we can rewrite the thermodynamical generating functional for the interacting theory (\ref{functional fourier}) as

\begin{align}
Z_{GF} =&\,\det_P\left(m_P\right)^{-3}\int_{P}\mathcal{D}A\mathcal{D}\overline{C}\mathcal{D}C\int_{A-P}\mathcal{D}\overline{\psi }\mathcal{D}\psi
e^{-S_T}\notag\\
\hspace{-.1cm}&\,\times\hspace{-.1cm}\exp \left[ \int_{\beta }d^{4}x\left( J_{\mu }A_{\mu }\hspace{-.1cm}+\overline{\eta }_{a}\psi _{a}\hspace{-.1cm}-\overline{\psi }_{a}\eta _{a}\hspace{-.1cm}-\overline{C}d+\overline{d}
C\right) \right] \hspace{-.1cm}.\label{aquatio}
\end{align}

From this thermodynamical generating functional we can compute any ensemble averages of the fields through, \textit{e.g.}, relations (\ref{leslie}) and (\ref{nielsen}). The Ward-Fradkin-Takahashi identities for the GQED$_4$ also follow from this expression.

\section{The Dyson-Schwinger-Fradkin equations}\label{DSF}

In this section we derive the Dyson-Schwinger-Fradkin equations for the GQED$_4$ in thermodynamic equilibrium. The Dyson-Schwinger-Fradkin equations comprise an infinite number of coupled nonlinear equations regarding all the Green functions of the theory \cite{Fradkin zero temperature, Dyson, Schwinger}.

First of all, we notice that the thermodynamical action (\ref{thermodynamical action}) is invariant under $\psi_a\rightarrow-\psi_a$ and $\overline{\psi}_a\rightarrow-\overline{\psi}_a$. Therefore, using the solution (\ref{aquatio}) for the thermodynamical generating functional, we find

\begin{align}
\left\langle\widehat{\psi}_a\left(\mathbf{x},\tau\right)\right\rangle=
\left\langle\widehat{\overline{\psi}}_a\left(\mathbf{x},\tau\right)\right\rangle=&\,0.\label{desta}
\end{align}

However, it is not possible to show that the thermal average of the Podolsky field is zero. We can only show it is an odd function of the parameter $q_e$. Nevertheless, from the technique employed in section \ref{Green}, we can show that such a quantity is point-independent:

\begin{align}
\left\langle\widehat{A}_\mu\left(\mathbf{x},\tau\right)\right\rangle=\left\langle\widehat{A}_\mu\right\rangle.\label{ateesta}
\end{align}

Now, we define the quantity:

\begin{align}
W\left[J,\eta ,\overline{\eta },d,\overline{d}\right]\equiv\ln\left\{Z_{GF}\left[ J,\eta ,\overline{\eta },d,\overline{d}\right]\right\}.
\end{align}

$W$ is called the thermodynamical generating functional of the \textit{connected} Green functions. We also define the derivatives of this generating functional as:

\begin{align}
\vartheta_\mu\left(\mathbf{x},\tau_x\right)\equiv&\,\frac{\delta W\left[J,\eta ,\overline{\eta },d,\overline{d}\right]}{\delta J_\mu\left(\mathbf{x},\tau_x\right)};\\
\overline{\chi}_a\left(\mathbf{x},\tau_x\right)\equiv&\,\frac{\delta W\left[J,\eta ,\overline{\eta },d,\overline{d}\right]}{\delta \eta_a\left(\mathbf{x},\tau_x\right)};\\
\chi_a\left(\mathbf{x},\tau_x\right)\equiv&\,\frac{\delta W\left[J,\eta ,\overline{\eta },d,\overline{d}\right]}{\delta \overline{\eta}_a\left(\mathbf{x},\tau_x\right)};\\
\mathcal{D}^{\left[s\right]}_{\mu\nu}\left(\mathbf{x},\mathbf{y};\tau_x,\tau_y\right)\equiv&\,\frac{\delta \vartheta_\mu\left(\mathbf{x},\tau_x\right)}{\delta J_\nu\left(\mathbf{y},\tau_y\right)};\label{mathD}\\
\mathcal{S}^{\left[s\right]}_{ab}\left(\mathbf{x},\mathbf{y};\tau_x,\tau_y\right)\equiv&\,\frac{\delta \chi_a\left(\mathbf{x},\tau_x\right)}{\delta\eta_b\left(\mathbf{y},\tau_y\right)}.\label{mathS}
\end{align}

Writing $Z_{GF}=e^W$ in the functional equations (\ref{leite}) and (\ref{pastel}), properly deriving these two equations and using these just defined quantities, we find

\begin{widetext}
\begin{align}
\delta_{\mu \nu }\delta \left( \mathbf{x}-\mathbf{y}\right) \delta \left( \tau
_{x}-\tau _{y}\right)=&\,P_{\mu \xi }^{\left( m_{P}^{2},\alpha \right) }\left( \mathbf{x},\tau_x\right) \mathcal{D}_{ \xi \nu}^{\left[ s\right] }\left( \mathbf{x},\mathbf{y};\tau _{x},\tau
_{y}\right)-iq_{e}\left( \gamma _{\mu }^{E}\right) _{ab}\frac{\delta \mathcal{S}_{ba}^{\left[ s\right] }\left( \mathbf{x},\mathbf{x};\tau _{x},\tau
_{x}\right) }{\delta J_{\nu }\left( \mathbf{y},\tau _{y}\right) } ;\label{pig}\\
\delta _{ab}\delta \left( \mathbf{x}-\mathbf{y}\right) \delta \left( \tau_{x}-\tau _{y}\right)=&\,\left\{ \left( \gamma _{\mu }^{E}\right) _{ac}D_{\mu }^{\left( \mu_{e},q_{e}\right) }\hspace{-.1cm}\left[ \frac{\delta }{\delta J\left( \mathbf{x},\tau
_{x}\right) }+\vartheta \left( \mathbf{x},\tau _{x}\right) \right]\hspace{-.1cm}
-m_{f}\delta _{ac}\right\} \mathcal{S}_{cb}^{\left[ s\right] }\left( \mathbf{x},\mathbf{y};\tau _{x},\tau _{y}\right) ;\label{pork}\\
J_{\mu }\left( \mathbf{x},\tau_{x}\right)=&\,P_{\mu \nu }^{\left( m_{P}^{2},\alpha \right)
}\left( \mathbf{x},\tau_x\right) \vartheta _{\nu }\left( \mathbf{x},\tau _{x}\right)\notag\\
&\,-iq_{e}\left( \gamma _{\mu }^{E}\right) _{ab}\left[ \mathcal{S}_{ba}^{\left[ s\right] }\left( \mathbf{x},\mathbf{x};\tau _{x},\tau _{x}\right) -\chi _{b}\left(\mathbf{x},\tau _{x}\right)\overline{\chi }_{a}\left( \mathbf{x},\tau _{x}\right)\right] ;\\
\eta _{a}\left( \mathbf{x},\tau _{x}\right)=&\,\left\{ \left( \gamma _{\mu }^{E}\right) _{ab}D_{\mu }^{\left( \mu_{e},q_{e}\right) }\left[ \frac{\delta }{\delta J\left( \mathbf{x},\tau
_{x}\right) }+\vartheta \left( \mathbf{x},\tau _{x}\right) \right]
-m_{f}\delta _{ab}\right\} \chi_{b}\left( \mathbf{x},\tau _{x}\right)  .
\end{align}
\end{widetext}

Now we define the polarization $\Pi_{\mu\nu}^{\left[s\right]}$ and the mass $\Sigma_{ab}^{\left[s\right]}$ operators implicitly through the following relations:

\begin{widetext}
\begin{align}
iq_{e}\left( \gamma _{\mu }^{E}\right) _{ab}\frac{\delta \mathcal{S}_{ba}^{\left[ s\right] }\left( \mathbf{x},\mathbf{x};\tau _{x},\tau
_{x}\right) }{\delta J_{\nu }\left( \mathbf{y},\tau _{y}\right) }\equiv&\,-\int_\beta d^4z\Pi_{\mu\xi}^{\left[s\right]}\left(\mathbf{x},\mathbf{z};\tau_x,\tau_z\right) \mathcal{D}_{\xi\nu}^{\left[s\right]}\left(\mathbf{z},\mathbf{y};\tau_z,\tau_y\right);\\
iq_{e}\left( \gamma _{\mu }^{E}\right) _{ab}\frac{\delta \mathcal{S}_{cb}^{\left[ s\right] }\left( \mathbf{x},\mathbf{y};\tau _{x},\tau
_{y}\right) }{\delta J_{\mu }\left( \mathbf{x},\tau _{x}\right) }\equiv&\,-\int_\beta d^4z\Sigma_{ac}^{\left[s\right]}\left(\mathbf{x},\mathbf{z};\tau_x,\tau_z\right) \mathcal{S}_{cb}^{\left[s\right]}\left(\mathbf{z},\mathbf{y};\tau_z,\tau_y\right)
\end{align}
\end{widetext}

With the aid of these two operators, equations (\ref{pig}) and (\ref{pork}) show that the quantities (\ref{mathD}) and (\ref{mathS}) are the Green functions of the following \textit{complete} operators:

\begin{widetext}
\begin{align}
\left[\mathcal{D}_{\mu\nu}^{\left[s\right]}\left(\mathbf{x},\mathbf{y};\tau_x,\tau_y\right)\right]^{-1}=&\,\delta_{\mu \xi }\delta \left( \mathbf{x}-\mathbf{y}\right) \delta \left( \tau
_{x}-\tau _{y}\right)P_{\xi\nu }^{\left( m_{P}^{2},\alpha \right) }\left( y\right) + \Pi_{\mu\nu}^{\left[s\right]}\left(\mathbf{x},\mathbf{y};\tau_x,\tau_y\right);\label{fusili}\\
\left[\mathcal{S}_{ab}^{\left[s\right]}\left(\mathbf{x},\mathbf{y};\tau_x,\tau_y\right)\right]^{-1}=&\,\delta _{ac}\delta \left( \mathbf{x}-\mathbf{y}\right) \delta \left( \tau_{x}-\tau _{y}\right)\left\{\left( \gamma _{\mu }^{E}\right) _{cb}D_{\mu }^{\left( \mu_{e},q_{e}\right) }\left[\vartheta \left( \mathbf{y},\tau _{y}\right) \right]
-m_{f}\delta _{cb}\right\}+\notag\\
&\,-\Sigma_{ab}^{\left[s\right]}\left(\mathbf{x},\mathbf{y};\tau_x,\tau_y\right).\label{penne}
\end{align}
\end{widetext}

The reason why these operators are called ``complete" is because they take into account all the interactions of the quantum theory. It is worth to emphasize that the functions (\ref{mathD}) and (\ref{mathS}) are the Green functions of the operators (\ref{fusili}) and (\ref{penne}) \textit{regardless} the presence of the external sources.  From the very definitions of the polarization and mass operators we can prove the following relations

\begin{widetext}
\begin{align}
\Pi _{\mu \nu }^{\left[ s\right] }\left( \mathbf{x},\mathbf{y};\tau
_{x},\tau _{y}\right)  =&\,\left( q_{e}\right) ^{2}\left( \gamma _{\mu
}^{E}\right) _{ab}\int_{\beta }d^{4}ud^{4}v\,\mathcal{S}_{bc}^{\left[ s\right]
}\left( \mathbf{x},\mathbf{u};\tau _{x},\tau _{u}\right) \Gamma _{\nu \left(
cd\right) }^{\left[ s\right] }\left( \mathbf{u},\mathbf{v},\mathbf{y};\tau
_{u},\tau _{v},\tau _{y}\right)\notag\\
&\,\times \mathcal{S}_{da}^{\left[ s\right] }\left(
\mathbf{v},\mathbf{x};\tau _{v},\tau _{x}\right) ; \\
\Sigma _{ab}^{\left[ s\right] }\left( \mathbf{x},\mathbf{y};\tau _{x},\tau
_{y}\right)  =&\,-\left( q_{e}\right) ^{2}\left( \gamma _{\mu }^{E}\right)
_{ac}\int_{\beta }d^{4}ud^{4}v\,\mathcal{D}_{\mu \nu }^{\left[ s\right]
}\left( \mathbf{x},\mathbf{u};\tau _{x},\tau _{u}\right) \mathcal{S}_{cd}^{\left[ s\right] }\left( \mathbf{x},\mathbf{v};\tau _{x},\tau _{v}\right) \notag\\
&\,\times\Gamma _{\nu \left( db\right) }^{\left[ s\right] }\left( \mathbf{v},\mathbf{y},\mathbf{u};\tau _{v},\tau _{y},\tau _{u}\right).
\end{align}
\end{widetext}
where  $\Gamma ^{\left[ s\right] }_{\mu \left( ab\right) }$ is the \textit{complete vertex function}:

\begin{widetext}
\begin{align}
\Gamma _{\mu \left( ab\right) }^{\left[ s\right] }
\left( \mathbf{x},\mathbf{y},\mathbf{z};\tau _{x},\tau _{y},\tau _{z}\right) \equiv &\,-\frac{i}{q_{e}}\frac{\delta \left\{ \left[ \mathcal{S}_{ab}^{\left[ s\right] }\left(
\mathbf{x},\mathbf{y};\tau _{x},\tau _{y}\right) \right] ^{-1}\right\} }{\delta \vartheta _{\mu }\left( \mathbf{z},\tau _{z}\right) } \notag\\
=&\,\left( \gamma _{\mu }^{E}\right) _{ab}\delta \left( \mathbf{x}-\mathbf{y}\right) \delta \left( \mathbf{z}-\mathbf{y}\right) \delta \left( \tau
_{x}-\tau _{y}\right) \delta \left( \tau _{z}-\tau _{y}\right)\notag\\
 &\,+\frac{i}{q_{e}}\frac{\delta \Sigma _{ab}^{\left[ s\right] }\left( \mathbf{x},\mathbf{y%
};\tau _{x},\tau _{y}\right) }{\delta \vartheta _{\mu }\left( \mathbf{z},\tau _{z}\right) }.
\end{align}
\end{widetext}

We see that in the absence of interaction, $\Gamma _{\mu \left( ab\right) }\propto \left(\gamma_\mu^E\right)_{ab}$.

Computing the complete Green functions (\ref{mathD}) and (\ref{mathS}) for the case of vanishing sources and using (\ref{limbo}), (\ref{limboso}), (\ref{desta}), and (\ref{ateesta}), we find

\begin{align}
\mathcal{S}^{\left[0\right]}_{ab}\left(\mathbf{x},\mathbf{y};\tau_x,\tau_y\right)=&\,S_{ab}\left( \mathbf{x}-\mathbf{y},\tau _{x}-\tau_{y}\right)
\equiv\mathcal{S}_{ab}\left( \mathbf{x}-\mathbf{y},\tau _{x}-\tau_{y}\right);\\
\mathcal{D}^{\left[0\right]}_{\mu\nu}\left(\mathbf{x},\mathbf{y};\tau_x,\tau_y\right)=&\,D_{\mu \nu }\left( \mathbf{x}-\mathbf{y},\tau _{x}-\tau_{y}\right)-\left\langle\widehat{A}_\mu\right\rangle\left\langle\widehat{A}_\nu\right\rangle
\equiv\mathcal{D}_{\mu \nu }\left( \mathbf{x}-\mathbf{y},\tau _{x}-\tau_{y}\right).\label{D completo}
\end{align}

Therefore, in the absence of the classical sources, the polarization and the mass operators also depend only on the differences $ \mathbf{x}-\mathbf{y}$ and $\tau _{x}-\tau_{y}$. However, as we have called attention at the end of section (\ref{MF formalism}), the parameter $\tau$ of the functional derivatives is restricted to the interval $\left(0,\beta\right)$. Due to the periodicity of the function $\mathcal{D}_{\mu \nu }$, for consistence we shall define the functional derivatives outside the original interval, replacing every Dirac delta distribution dependent on the $\tau$ variable that appears in the equation (\ref{fusili}) by the \textit{periodic Dirac comb distribution} $\Delta^+_\beta\left(\tau\right)$. A similar statement is true for the delta distributions appearing in the equation for $\mathcal{S}_{ab}$, replacing them by the \textit{anti-periodic Dirac comb distribution} $\Delta^-_\beta\left(\tau\right)$. These distributions are defined as

\begin{equation}
\Delta^\pm_\beta\left(\tau\right)\equiv\sum_{n=-\infty}^\infty \left(\pm 1\right)^n\delta\left(\tau-n\beta\right).
\end{equation}

The Dirac combs satisfy the same periodicity conditions of their corresponding Green functions:

\begin{align}
\Delta^\pm_\beta\left(\tau-\beta\right)=\pm\Delta^\pm_\beta\left(\tau\right).
\end{align}

Now, we can show that the complete Green functions in thermodynamic equilibrium satisfy:

\begin{widetext}
\begin{align}
\delta_{\mu \nu }\delta \left( \mathbf{x}-\mathbf{y}\right) \Delta_\beta^+ \left( \tau
_{x}-\tau _{y}\right)=&\,\hspace{-.1cm}\int_\beta d^4z\hspace{-.1cm}\left[\delta_{\mu \xi }\delta \left( \mathbf{x}-\mathbf{z}\right) \Delta_\beta^+ \left( \tau
_{x}-\tau _{z}\right)P_{\xi\sigma }^{\left( m_{P}^{2},\alpha \right) }\left( z\right) \right.\notag\\
&\,\left.+ \Pi_{\mu\sigma}\left(\mathbf{x}-\mathbf{z},\tau_x-\tau_z\right)\right] \mathcal{D}_{\sigma \nu }\left( \mathbf{z}-\mathbf{y},\tau _{z}-\tau_{y}\right);\\
\delta _{ab}\delta \left( \mathbf{x}-\mathbf{y}\right) \Delta_\beta^- \left( \tau_{x}-\tau _{y}\right)=&\,\int_\beta d^4z\left\{\delta _{ac}\delta \left( \mathbf{x}-\mathbf{z}\right) \Delta_\beta^- \left( \tau_{x}-\tau _{z}\right)\left\{\left( \gamma _{\mu }^{E}\right) _{cd}D_{\mu }^{\left( \mu_{e},q_{e}\right) }\left[\left\langle\widehat{A}\right\rangle \right]\right.\right.\notag\\
&\,\left.\left.
-m_{f}\delta _{cd}\right\}_z-\Sigma_{ad}\left(\mathbf{x}-\mathbf{z},\tau_x-\tau_z\right)\right\}\mathcal{S}_{db}\left( \mathbf{z}-\mathbf{y},\tau _{z}-\tau_{y}\right).
\end{align}
\end{widetext}

 In order to clarify the role of the thermal equilibrium, we shall write these expressions in Fourier space. Since the quantities appearing in these expressions are either periodic or anti-periodic, the Fourier variable associated with $\tau$ can only assume discrete values:

 \begin{align}
 \mathcal{D}_{\mu \nu }\left( \mathbf{x},\tau _{x}\right)  =&\,\sum_{n=-\infty}^{+\infty}\int \frac{d^{3}k}{\beta \left( 2\pi \right) ^{3}}
\tilde{\mathcal{D}}_{\mu \nu }\left( \mathbf{k},\omega _{n}^{B}\right) e^{i\left(
\omega _{n}^{B} \tau _{x} +\mathbf{k}\cdot \mathbf{x}\right) };\\
\mathcal{S}_{ab }\left( \mathbf{x},\tau _{x}\right)  =&\,\sum_{n=-\infty}^{+\infty}\int \frac{d^{3}k}{\beta \left( 2\pi \right) ^{3}}
\tilde{\mathcal{S}}_{ab }\left( \mathbf{k},\omega _{n}^{F}\right) e^{i\left(
\omega _{n}^{F} \tau _{x} +\mathbf{k}\cdot \mathbf{x}\right) };\\
\Delta_\beta^+ \left( \tau\right)  =&\,\frac{1}{\beta}\sum_{n=-\infty}^{+\infty}e^{i\omega _{n}^{B} \tau _{x}};\\
\Delta_\beta^- \left( \tau\right)  =&\,\frac{1}{\beta}\sum_{n=-\infty}^{+\infty}e^{i\omega _{n}^{F} \tau _{x}}.
 \end{align}

Here, we have defined both the \textit{bosonic} and the \textit{fermionic Matsubara frequencies} respectively as $\omega _{n}^{B}\equiv2n\pi/\beta$ and $\omega_n^F\equiv\left(2n+1\right)\pi/\beta$, for $n\in \mathbb{Z}$. With similar Fourier expansions for the polarization and mass operators, we find

\begin{widetext}
\begin{align}
\left[ \widetilde{P}_{\mu \xi }^{\left( m_{P}^{2},\alpha \right) }\left(k^{Bn}\right) +\tilde{\Pi} _{\mu \xi }\left( \mathbf{k},\omega
_{n}^{B}\right) \right] \tilde{\mathcal{D}}_{\xi \nu }\left( \mathbf{k},\omega
_{n}^{B}\right) =&\,\delta _{\mu \nu };\label{quase}\\
\left[ i\left( \gamma _{\mu }^{E}\right) _{ac}\left( k_{\mu }^{F_{n}}-i\mu
_{e}\delta _{\mu 0}+q_{e}\left\langle \widehat{A}_{\mu }\right\rangle
\right) -m_{f}\delta _{ac}-\tilde{\Sigma} _{ac}\left( \mathbf{k},\omega
_{n}^{F}\right) \right] \tilde{\mathcal{S}}_{cb}\left( \mathbf{k},\omega
_{n}^{F}\right) =&\,\delta _{ab}.\label{chorona}
\end{align}
\end{widetext}

In these expressions, $k^{Bn} _{\mu }\equiv\left(\omega_n^B,\mathbf{k}\right)$, $k^{Fn} _{\mu }\equiv\left(\omega_n^F,\mathbf{k}\right)$, and $\widetilde{P}_{\mu \nu }^{\left( m_{P}^{2},\alpha \right) }$ is the Fourier transform of the Podolsky differential operator (\ref{Podolsky differential}):

\begin{align}
\widetilde{P}_{\mu \nu }^{\left( m_{P}^{2},\alpha \right) }\left(k^{Bn}\right) \equiv&\,-\left[ \frac{\left(k^{Bn}\right)^2 }{m_{P}^{2}}+1\right] \left\{ \left(k^{Bn}\right)^2  \delta _{\mu \nu }-\left[ 1-\frac{1}{\alpha }\left( \frac{\left(k^{Bn}\right)^2}{m_{P}^{2}}+1\right) \right]
k^{Bn} _{\mu }k^{Bn} _{\nu }\right\},\label{roma}
\end{align}
where  $\left(k^{Bn}\right)^2\equiv k^{Bn} _{\mu }k^{Bn} _{\mu }=\left(\omega_n^B\right)^2+\mathbf{k}^2\geq 0$.

We notice that the use of the Dirac comb distributions allowed us to write equations for the complete Green functions of the theory in a local fashion in the Fourier space. Furthermore, these equations resemble the corresponding expressions at zero temperature. The main differences appear in (\ref{chorona}). This equation  depends explicitly on both the chemical potential $\mu_e$  and the thermal average of the Podolsky operator $\left\langle \widehat{A}_{\mu }\right\rangle$.

For completeness, we write the equation for the ghost Green function in the Fourier space:

\begin{equation}
\frac{i}{\sqrt{\alpha}}\left[\frac{\left(k^{Bn}\right)^2}{m_P^2}+1\right]\left(k^{Bn}\right)^2\tilde{G}\left(\omega_n^B,\mathbf{k}\right)=1.
\end{equation}

Despite the ghost fields be Grassmannian, we have fixed the ghost chemical potential (\ref{muzinho}) to be of the form $\mu_g=i\omega_n^F$, which implies a periodicity condition over the ghost Green function (\ref{bacteria}). If $G$ is periodic, its Fourier transform must depend on \textit{bosonic} Matsubara frequencies.

\section{The Ward-Fradkin-Takahashi Identities}\label{WFTsection}

The Ward-Fradkin-Takahashi are relations among the (complete) Green functions of the theory \cite{Ward, FradkinWFT, Takahashi}. In the present section we shall demonstrate them for the GQED$_4$ in thermodynamic equilibrium.

Let us start by changing the integration functions in the path integral representation for the thermodynamical generating functional (\ref{aquatio}) accordingly to:

\begin{eqnarray}
A_{\mu }\left( \mathbf{x},\tau \right)  &\rightarrow &A_{\mu }^{\prime
}\left( \mathbf{x},\tau \right) =A_{\mu }\left( \mathbf{x},\tau \right)
-\partial _{\mu }\zeta \left( \mathbf{x},\tau \right) ; \\
\psi _{a}\left( \mathbf{x},\tau \right)  &\rightarrow &\psi _{a}^{\prime
}\left( \mathbf{x},\tau \right) =e^{iq_{e}\zeta \left( \mathbf{x},\tau
\right)}\psi _{a}\left( \mathbf{x},\tau \right) ; \\
\overline{\psi }_{a}\left( \mathbf{x},\tau \right)  &\rightarrow &\overline{%
\psi }_{a}^{\prime }\left( \mathbf{x},\tau \right) =\overline{\psi }_{a}\left( \mathbf{x},\tau \right)e^{-iq_{e}\zeta \left( \mathbf{x},\tau
\right)};  \\
C\left( \mathbf{x},\tau \right)  &\rightarrow &C^{\prime }\left( \mathbf{x}%
,\tau \right) =C\left( \mathbf{x},\tau \right) ; \\
\overline{C}\left( \mathbf{x},\tau \right)  &\rightarrow &\overline{C}%
^{\prime }\left( \mathbf{x},\tau \right) =\overline{C}\left( \mathbf{x},\tau
\right) ,
\end{eqnarray}

Here, $ \zeta\left( \mathbf{x},\tau\right)$ is a real, periodic function. The generating functional's integration measure $\mathcal{D}A\mathcal{D}\overline{C}\mathcal{D}C\mathcal{D}\overline{\psi }\mathcal{D}\psi$ is invariant under this $U(1)$ transformation \cite{Chaichian}. Therefore, the thermodynamical generating functional can be written as

\begin{align}
Z_{GF} =&\,\det_{P}\left( m_{P}\right) ^{-3}\int_{P}\mathcal{D}A\mathcal{D}\overline{%
C}\mathcal{D}C\int_{A-P}\mathcal{D}\overline{\psi }\mathcal{D}\psi
e^{-S_{T}}  \notag \\
&\,\times\exp\left\{- \int_{\beta }d^{4}y\,\left\{\zeta \left[ \frac{1}{%
\alpha }\left( \frac{\Delta }{m_{P}^{2}}+1\right) ^{2}\Delta \partial _{\nu
}A_{\nu }\right.\right.+\partial _{\nu }J_{\nu }\right]\notag\\
&\,\left.\left.-\frac{1}{2}\partial_\mu\zeta P_{\mu\nu}^{\left(m_P^2,\alpha\right)}\partial_\nu\zeta
-\overline{\eta}_a\left(e^{iq_e\zeta}-1\right)\psi_a-\overline{\psi}_a\left(e^{-iq_e\zeta}-1\right)\eta_a\right\}\right\}\notag\\
&\,\times \hspace{-.1cm}\exp \left[ \int_{\beta }\hspace{-.1cm}d^{4}x\left( J_{\mu }A_{\mu }\hspace{-.1cm}+\overline{%
\eta }_{a}\psi _{a}\hspace{-.1cm}-\overline{\psi }_{a}\eta _{a}\hspace{-.1cm}-\overline{C}d+\overline{d}C%
\right) \right].\label{delicia}
\end{align}

Since $Z_{GF}$ in equation (\ref{aquatio}) is originally independent of the function $ \zeta\left( \mathbf{x},\tau\right)$, the following relation holds:

\begin{equation}
\left. \frac{\delta Z_{GF}}{\delta \zeta \left( \mathbf{x},\tau
_{x}\right) }\right\vert _{\zeta =0}=0.
\end{equation}

This condition applied to (\ref{delicia}) implies:

\begin{align}
\partial _{\mu }\frac{\delta \Gamma }{\delta \vartheta _{\mu }\left( \mathbf{%
x},\tau _{x}\right) } =&\,\frac{1}{\alpha }\left( \frac{\Delta }{m_{P}^{2}}%
+1\right) ^{2}\Delta \partial _{\mu }\vartheta _{\mu }\left( \mathbf{x},\tau
_{x}\right) \notag\\
&\,+iq_{e}\left[ \frac{\delta \Gamma }{\delta \chi _{a}\left( \mathbf{x},\tau
_{x}\right) }\chi _{a}\left( \mathbf{x},\tau _{x}\right) -\frac{\delta
\Gamma }{\delta \overline{\chi }_{a}\left( \mathbf{x},\tau _{x}\right) }%
\overline{\chi }_{a}\left( \mathbf{x},\tau _{x}\right) \right] .\label{WFT}
\end{align}

The function $\Gamma$ that appears in this expression is the Legendre transform of the functional $W$:\footnote{We could have defined $\Gamma$ as the Legendre transform of the ghost fields as well. However, since they play no role in the Ward-Fradkin-Takahashi identities for the Podolsky theory, we have opted for the definition (\ref{azul}).}

\begin{align}
\Gamma \left[ \vartheta ,\chi ,\overline{\chi };d,\overline{d}\right] \equiv&\,
W\left[ J,\eta ,\overline{\eta },d,\overline{d}\right] -\int_{\beta }d^{4}x%
\left[ J_{\mu }\vartheta _{\mu }+\overline{\eta }_{a}\chi _{a}-\overline{%
\chi }_{a}\eta _{a}\right] .\label{azul}
\end{align}

By deriving this definition we can show:

\begin{align}
\frac{\delta ^{2}\Gamma }{\delta \chi _{b}\left( \mathbf{y},\tau _{y}\right)
\delta \overline{\chi }_{a}\left( \mathbf{x},\tau _{x}\right) }=&\,\left[
\mathcal{S}_{ab}^{\left[ s\right] }\left( \mathbf{x},\mathbf{y};\tau
_{x},\tau _{y}\right) \right] ^{-1};\\
\frac{\delta ^{2}\Gamma }{\delta \vartheta _{\nu }\left( \mathbf{y},\tau
_{y}\right) \delta \vartheta _{\mu }\left( \mathbf{x},\tau _{x}\right) }=&\,-
\left[ \mathcal{D}_{\mu \nu }^{\left[ s\right] }\left( \mathbf{x},\mathbf{y}%
;\tau _{x},\tau _{y}\right) \right] ^{-1}.
\end{align}

From (\ref{WFT}) an infinite number of relations, known as Ward-Fradkin-Takahashi \cite{Ward, FradkinWFT, Takahashi}, follows. For instance, by properly deriving that relation and setting the external sources to vanish, we get the following identities:

\begin{widetext}
\begin{align}
\partial _{\mu }^{\left( x\right) }\Gamma _{\mu \left( ab\right) }\left(
\mathbf{z},\mathbf{y},\mathbf{x};\tau _{z},\tau _{y},\tau _{x}\right) =&\,
\left[ \mathcal{S}_{ab}\left( \mathbf{x}-\mathbf{z},\tau _{x}-\tau
_{z}\right) \right] ^{-1}\delta \left( \mathbf{x}-\mathbf{y}\right) \Delta
_{\beta }^{-}\left( \tau _{x}-\tau _{y}\right)  \notag\\
&\,-\left[ \mathcal{S}_{ab}\left( \mathbf{y}-\mathbf{x},\tau _{y}-\tau
_{x}\right) \right] ^{-1}\delta \left( \mathbf{x}-\mathbf{z}\right) \Delta
_{\beta }^{-}\left( \tau _{x}-\tau _{z}\right);\label{lobisomem}\\
\partial _{\mu }\left[ \mathcal{D}_{\mu \nu }\left( \mathbf{x}-\mathbf{y}%
,\tau _{x}-\tau _{y}\right) \right] ^{-1}=&\,-\frac{1}{\alpha }\left( \frac{%
\Delta }{m_{P}^{2}}+1\right) ^{2}\Delta \partial _{\nu }\delta \left(
\mathbf{x}-\mathbf{y}\right) \Delta _{\beta }^{+}\left( \tau _{x}-\tau
_{y}\right) .\label{vampiro}
\end{align}
\end{widetext}

By writing equation (\ref{lobisomem}) in Fourier space we get the \textit{Ward identity} in thermodynamic equilibrium
\cite{Ward}:

\begin{align}
p_{\mu }^{Bl}\widetilde{\Gamma }_{\mu \left( ab\right) }\left( \mathbf{k},\omega _{n}^{F};\mathbf{p},\omega _{l}^{B}\right) =&\,\widetilde{\mathcal{S}}
_{ab}^{-1}\left( \mathbf{k}+\mathbf{p},\omega _{n+l}^{F}\right)-\widetilde{\mathcal{S}}_{ab}^{-1}\left( \mathbf{k},\omega _{n}^{F}\right) ,\label{minhon}
\end{align}

By Fourier-transforming equation (\ref{vampiro}) we find:

\begin{equation}
k_{\mu }^{Bn}\widetilde{\mathcal{D}}_{\mu \nu }^{-1}\left( \mathbf{k},\omega
_{n}^{B}\right) =-\frac{1}{\alpha }\left[ \frac{\left( k^{Bn}\right) ^{2}}{%
m_{P}^{2}}+1\right] ^{2}\left( k^{Bn}\right) ^{2}k_{\nu }^{Bn}.
\end{equation}

Since the right-hand-side of this equation does not depend on the coupling constant $q_e$, this relation holds also in the free case. This can be explicitly checked by computing $k_{\mu }^{Bn}\widetilde{P}_{\mu \nu }^{\left( m_{P}^{2},\alpha \right) }\left(k^{Bn}\right)$ from equation (\ref{roma}). From equation (\ref{quase}) we identify the Fourier transform of the polarization operator as $\widetilde{\Pi}_{\mu\nu}\left(\mathbf{k},\omega_n^B\right)=\widetilde{\mathcal{D}}_{\mu \nu }^{-1}\left( \mathbf{k},\omega_{n}^{B}\right) -\widetilde{P}_{\mu \nu }^{\left( m_{P}^{2},\alpha \right) }\left(k^{Bn}\right)$ and, therefore, we have proved the \textit{transversality of the polarization operator} \cite{FradkinWFT}:

\begin{equation}
k_\mu^{Bn}\widetilde{\Pi}_{\mu\nu}\left(\mathbf{k},\omega_n^B\right)=0.
\end{equation}

It is important to emphasize that the transversality of $\Pi_{\mu\nu}$ holds also at finite temperature, despite the presence of the Podolsky mass term. The central issue here is that the Podolsky theory is gauge invariant, and this symmetry is not broken by thermal effects.

Using equations (\ref{limbo}) and (\ref{D completo}), we can show the symmetry of the Podolsky Green function: $\mathcal{D}_{\mu\nu}\left(\mathbf{x},\tau\right)=\mathcal{D}_{\nu\mu}\left(-\mathbf{x},-\tau\right)$. This implies the symmetry of the polarization: $\widetilde{\Pi}_{\mu\nu}\left(\mathbf{k},\omega_n^B\right)=\widetilde{\Pi}_{\nu\mu}\left(-\mathbf{k},-\omega_n^B\right)$. From this and from the transversality condition, we can write the most general form for the polarization operator for the GQED$_4$ in thermodynamic equilibrium:

\begin{widetext}
\begin{align}
\widetilde{\Pi }_{\mu \nu }\left( \mathbf{k},\omega _{n}^{B}\right)
=&\,A\left( \mathbf{k},\omega _{n}^{B}\right) \left[ \delta _{\mu
\nu }-\frac{k_{\mu }^{Bn}k_{\nu }^{Bn}}{\left( k^{Bn}\right) ^{2}}\right]\notag\\
&\,+B\left( \mathbf{k},\omega _{n}^{B}\right) \left[ \frac{k_{\mu
}^{Bn}k_{\nu }^{Bn}}{\left( k^{Bn}\right) ^{2}}-\left( \frac{k_{\mu
}^{Bn}\delta _{\nu 0}+k_{\nu }^{Bn}\delta _{\mu 0}}{\omega _{n}^{B}}%
\right) +\frac{\left( k^{Bn}\right) ^{2}\delta _{\mu 0}\delta _{\nu 0}}{%
\left( \omega _{n}^{B}\right) ^{2}}\right] ,
\end{align}
\end{widetext}
where $A\left( \mathbf{k},\omega _{n}^{B}\right)=A\left(- \mathbf{k},-\omega _{n}^{B}\right)$ and $B\left( \mathbf{k},\omega _{n}^{B}\right)=B\left(- \mathbf{k},-\omega _{n}^{B}\right)$. The reason for this unusual form is that an interacting theory in thermodynamic equilibrium comprises a medium. In the present case, this medium is a \textit{plasma}. So, besides the usual momentum $k_\mu^{Bn}$, there is also the velocity $u_\mu$ of the medium available for building tensors up. In the plasma rest frame, the velocity $u_\mu$ is proportional to $\delta_{\mu 0}$, which explains the dependence of the polarization tensor on those quantities. The polarization tensor has the same form in Maxwell and Podolsky theories \cite{Fradkin relations in SM}. The only differences are the functions $A\left( \mathbf{k},\omega _{n}^{B}\right)$ and $B\left( \mathbf{k},\omega _{n}^{B}\right)$ where, in the case of the GQED$_4$, all the dependence on the Podolsky parameter lies.

\section{Final remarks}\label{final remarks}

Starting from the very concept of density matrix, we have reviewed the Matsubara-Fradkin formalism. As we have seen, there is a natural correspondence between the field dependence on the temperature-associated variable, denoted by $\tau$, through the similarity transformation of the field with the partition function with the time evolution of fields at zero temperature. Fundamentally, it is precisely this correspondence that justifies the analogy between the partition function with the transition amplitude found in the imaginary-time formalism. Besides, in this formalism, the Euclidean structure of the spacetime emerges naturally. In other words, we do not need to perform a Wick rotation as is usual in other techniques. Once we have finished the review of the method, we applied it to the thermodynamical quantization of the GQED$_4$.

We have shown that the fermionic part of the electrodynamical theory can be quantized \textit{a la} Dirac without further complications. Since this part of the theory does not depend on the explicit form of the Lagrangian density of the gauge field, the quantization presented here applies to both Maxwell and Podolsky theories. However, for the gauge part of the theory, we have opted for the Nakanishi's covariant quantization. The reason is that even if we had done otherwise, we would have had to pass from a non-covariant gauge choice to a covariant one, pretty much as we have done in our previous work in the imaginary time formalism \cite{Bonin}. We have also seen how the ghost fields arise in the Nakanishi's method and their relation with the residual gauge symmetry of the theory. From the field equations for every field we have written the set of functional equations satisfied by the thermodynamical generating functional. In order to do that, we have made use of the expressions of thermal averages of operators in terms of functional derivatives of the thermodynamical generating functional. We have proved the periodicity properties of the ordering of the Podolsky and of the fermionic fields in thermodyanmic equilibrium. The periodicity of the ghost fields, however, was initially depending on the value of the ghost chemical potential, which we showed it must be a pure imaginary number and, therefore, not a thermodynamical observable. Then, we have found path integral representations for the thermodynamical generating functional and for the partition function. Upon computing the partition function of the free Podolsky theory, we were able to choose an appropriate value for the ghost chemical potential. The correct value for the ghost chemical potential leads to a well-defined periodicity property for the ghost Green function. In computing the partition function for the free Podolsky field, we have rederived our previous result using a complete different and independent method \cite{Bonin}. It is important to stress that, in the present formalism, the path integral representation for the partition function was found as a solution to the set of functional equations satisfied by the thermodynamical generating functional. In other words, we did not need to call upon any kind of analogy to construct the correct partition function of the theory. We have also written the Dyson-Schwinger-Fradkin equations for the GQED$_4$. Furthermore we have written the Ward-Fradkin-Takahashi identities for the GQED$_4$ in thermodynamic equilibrium. From these identities we have proved the Ward identity and the transversality of the Podolsky polarization tensor. Using this result, we have written the most general form for the polarization tensor of the GQED$_4$ in thermodynamic equilibrium. Since there is a new four-vector available, namely, the four-velocity of the medium, this general form differs greatly from its zero temperature version.

We call attention to the following property of our results: if we take the limit $\beta\rightarrow \infty$ and $\mu\rightarrow 0$ (which corresponds to the vanishing of temperature and chemical potential) in any relation regarding a Green function in thermal equilibrium, we recover the corresponding relation in the zero-temperature Euclidean theory. This also holds in the Fourier space. The Matsubara frequencies, being them bosonic or fermionic, are inversely proportional to $\beta$. In the limit of vanishing temperature, the Matsubara frequencies become dense in the real numbers set: they become continuous variables and the sums over them become integrals. Furthermore, the Green functions in thermal equilibrium are either periodic or anti-periodic with period $\beta$. As this quantity goes to infinity, the thermal Green functions lose their periodicity properties and they become usual Euclidean Green functions. As a corollary, if we take these limits in any expression involving a Green function, like the Dyson-Schwinger-Fradkin and the Ward-Fradkin-Takahashi identities, we find Euclidean versions of those found in \cite{BPZ}. Besides, since we have not appealed to perturbative techniques, all our results hold in both perturbative and non-perturbative regimes of the GQED$_4$.

Now that we have set the basis for the thermodynamical quantization of the GQED$_4$, we can study plasma effects in the GQED$_4$, one of those, for instance, being the \textit{Debye screening}. In the Maxwell theory, the Debye screening consists of the Coulomb electrostatic potential becoming short ranged due to the interaction of the field with the plasma. In the GQED$_4$ the situation should be even more interesting, since we do not have a simple Coulomb potential, but a electrostatic potential a little bit more complicated. Plasma oscillations and the fermionic thermal mass generation should also be interesting problems in the GQED$_4$. Besides, the Matsubara-Fradkin thermodynamical quantization can be applied, in principle, to even more general interactions, like those of non-Abelian gauge theories.

\section{Acknowledgements}\label{acknowledgements}

C. A. B.  thanks Con\-se\-lho Na\-cio\-nal de De\-sen\-vol\-vi\-men\-to Cien\-t\'{i}\-fi\-co e Tec\-no\-l\'{o}\-gi\-co (CNPq) for total support. B. M. P. thanks both CNPq and Coordena\c{c}\~{a}o de Aperfei\c{c}oamento de Pessoal de N\'{i}vel Superior (CAPES) for partial support.

\end{document}